\documentclass[useAMS,usenatbib]{mn2e}
\usepackage{epsfig}
\usepackage{amssymb}
\usepackage{amsmath}
\usepackage{lscape}
\usepackage{longtable}
\usepackage[normal]{caption}
\usepackage{appendix}
\usepackage{bm}

\title[The star-formation rates of QSOs]{The star-formation rates of QSOs}

\author[M.~Symeonidis et al.] 
{\parbox{\textwidth}{\raggedright
M.~Symeonidis,$^{1}$\thanks{E-mail: \texttt{m.symeonidis@ucl.ac.uk}}
N. ~Maddox,$^{2,3}$
M. J. ~Jarvis,$^{4,5}$ 
M. J. ~Micha{\l}owski,$^{6}$ 
P. ~Andreani,$^{7}$
D. L. ~Clements,$^{8}$
G. ~De Zotti,$^{9}$
S. ~Duivenvoorden,$^{14}$
J. ~Gonzalez-Nuevo,$^{10,11}$
E. ~Ibar,$^{12}$
R. J. ~Ivison, $^{7}$
L. ~Leeuw,$^{13}$
M. J. ~Page,$^{1}$ 
R. ~Shirley,$^{14, 15, 16}$
M. W. L. ~Smith,$^{17}$
and M. ~Vaccari$^{18, 19}$
}\vspace{0.4cm}\\
\parbox{\textwidth}{\raggedright $^{1}$ Mullard Space Science
  Laboratory, University College London, Holmbury St. Mary, Dorking,
  Surrey RH5 6NT, UK\\
$^{2}$ The Netherlands Institute for Radio Astronomy, Oude Hoogeveensedijk 4, 7991 PD, Dwingeloo, The Netherlands\\
$^{3}$ University Observatory, Faculty of Physics, Ludwig-Maximilians-Universit\"at, Scheinerstr. 1, 81679 Munich, Germany\\
$^{4}$ University of Oxford, Denys Wilkinson Building, Keble Road, Oxford, OX1 3RH, UK\\
$^{5}$ Department for Physics, University of the Western Cape, Bellville 7535, South Africa\\
$^{6}$ Astronomical Observatory Institute, Faculty of Physics, Adam Mickiewicz University, ul.~S{\l}oneczna 36, 60-286 Pozna{\'n}, Poland\\
$^{7}$  European Southern Observatory, Karl Schwarzschild Strasse 2, D-85748 Garching, Germany\\
$^{8}$ Imperial College London, Prince Consort Road, London SW7 2AZ, UK\\
$^{9}$ INAF-Osservatorio Astronomico di Padova, Vicolo dell?Osservatorio 5, I-35122 Padova, Italy\\
$^{10}$ Departamento de Fisica, Universidad de Oviedo, C. Federico Garcia Lorca 18, 33007 Oviedo, Spain\\
$^{11}$ Instituto Universitario de Ciencias y Tecnologías Espaciales de Asturias (ICTEA), C. Independencia 13, 33004 Oviedo, Spain\\
$^{12}$ Instituto de F\'isica y Astronom\'ia, Universidad de Valpara\'iso, Avda. Gran Breta\~na 1111, Valpara\'iso, Chile \\
$^{13}$College of Graduate Studies, Barney Pityana Building, Room 01-010, Unisa Florida Science Campus, PO Box 392, UNISA, 0003, SOUTH AFRICA\\
$^{14}$Astronomy Centre, Department of Physics and Astronomy, University of Sussex, Brighton BN1 9QH, UK \\
$^{15}$Instituto de Astrof\'isica de Canarias, E-38205 La Laguna, Tenerife, Spain\\
$^{16}$Dpto. Astrof\'isica, Universidad de La Laguna, E-38206 La Laguna, Tenerife, Spain\\
$^{17}$ School of Physics and Astronomy, Cardiff University, Queens Buildings, The Parade, Cardiff, CF24 3AA, UK\\
$^{18}$ Department of Physics and Astronomy, University of the Western Cape, Robert Sobukwe Road, 7535 Bellville, Cape Town, South Africa\\
$^{19}$ INAF - Istituto di Radioastronomia, via Gobetti 101, 40129 Bologna, Italy\\
}}

\begin{document}

\date{Accepted  Received; in original form}

\pagerange{\pageref{firstpage}--\pageref{lastpage}} \pubyear{2014}

\maketitle

\label{firstpage}

\begin{abstract}

\noindent We examine the far-IR properties of a sample of 5391 optically selected QSOs in the $0.5<z<2.65$ redshift range down to log\,$[\nu L_{\nu,2500} (erg/s)]>44.7$, using SPIRE data from \textit{Herschel}-ATLAS. We split the sample in a grid of 74 luminosity-redshift bins and compute the average optical--infrared spectral energy distribution (SED) in each bin. By normalising an intrinsic AGN template to the AGN optical power (at 5100\AA) we decompose the total infrared emission ($L_{\rm IR}$; 8---1000$\mu$m) into an AGN ($L_{\rm IR, AGN}$) and star-forming component ($L_{\rm IR, SF}$). We find that the AGN contribution to $L_{\rm IR}$ increases as a function of AGN power which manifests as a reduction of the `far-IR bump' in the average QSO SEDs. We note that $L_{\rm IR, SF}$ does not correlate with AGN power; the mean star formation rates (SFRs) of AGN host galaxies are a function of redshift \textit{only} and they range from $\sim$6\,M$_{\odot}$/yr at $z\sim0$ to a plateau of $\lesssim$ 200\,M$_{\odot}$/yr at $z\sim2.6$. Our results indicate that the accuracy of far-IR emission as a proxy for SFR \textit{decreases} with increasing AGN luminosity. We show that, at any given redshift, observed trends between infrared luminosity (whether monochromatic or total) and AGN power (in the optical or X-rays) can be explained by a simple model which is the sum of two components: (A) the infrared emission from star-formation, uncorrelated with AGN power and (B) the infrared emission from AGN, directly proportional to AGN power in the optical or X-rays. 
\end{abstract}


\section{Introduction}
\label{sec:introduction}
Several observed phenomena and physical conditions point towards a
connection between nuclear activity and star
formation over the history of
both stellar and black hole mass build-up.
For example, the peak of QSO activity and the
peak of star formation are broadly coincident at $1<z<3$ (e.g. Boyle $\&$ Terlevich 1998\nocite{BT98}). Moreover there is an observed relationship between the mass of
the central black hole (BH) and the mass of the host galaxy bulge or stellar velocity dispersion ($\sigma$), often referred to as the `$M-\sigma$\
relation' (e.g. Magorrian et al. 1998\nocite{Magorrian98}, Ferrarese $\&$ Merritt 2000\nocite{FM00}, Gebhardt et al. 2000\nocite{Gebhardt00}). 
Such observations have given rise to the idea that there should be a connection between star-formation rate and AGN accretion rate, particularly because they are thought to draw from the same reservoir of fuel.  
Numerous studies have looked for this link, but the results have not converged. For example some studies appear to show that the AGN accretion rate correlates with star formation rate (SFR; e.g. Bonfield et al. 2011\nocite{Bonfield11}; Rovilos et al. 2012\nocite{Rovilos12}; Chen et al. 2013\nocite{Chen13}; Hickox et al. 2014\nocite{Hickox14}; Harris et al. 2016\nocite{Harris16}), whereas others report that there is no correlation at low AGN luminosities but there is one at high AGN luminosities (e.g. Shao et al. 2010\nocite{Shao10}; Rosario et al. 2012\nocite{Rosario12}, 2013\nocite{Rosario13}; Kalfountzou et al. 2012\nocite{Kalfountzou12}; Stanley et al. 2017\nocite{Stanley17}). Other studies report that the two quantities are not correlated (e.g. Rawlings et al. 2013\nocite{Rawlings13}; Azadi et al. 2015\nocite{Azadi15}; Stanley et al. 2015\nocite{Stanley15}; Pitchford et al. 2016\nocite{Pitchford16}; Schulze et al. 2019\nocite{Schulze19}; Bianchini et al. 2019\nocite{Bianchini19}). 
Complicating the extraction of correlations between nuclear 
luminosity and star-formation are the lack of sufficient spatial resolution, but also a number of underlying relations between stellar mass, black hole mass, SFR and AGN luminosity, as well as AGN variability (e.g. Hickox et al. 2014\nocite{Hickox14}).

Another issue affecting the determination of a correlation between
accretion and star formation is the difficulty in identifying an observational window that is either
dominated by one of the two processes, or where the contribution of
flux from each process can be accurately disentangled. For unobscured,
luminous QSOs, optical and ultraviolet (UV) wavelengths provide a
nearly uncontaminated measure of BH accretion, as the QSO outshines
the host galaxy by orders of magnitude. On the other hand, at longer wavelengths,
emission from processes within the host galaxy becomes increasingly important. While at mid-infrared (MIR)
wavelengths, emission is dominated by hot dust within the AGN torus (e.g. Osterbrock $\&$ Ferland 2006\nocite{OF06}; Rodriguez-Ardila $\&$ Mazzalay 2006\nocite{RAM06}), the primary cause for emission at far-infrared (FIR)
wavelengths ($>$60$\mu$m) has been the subject of much
debate. Due to the perceived confinement of the AGN emission to the immediate nuclear
environment, it is often assumed that star formation is the primary
source of heating for cool dust, with little contamination from the
AGN at $\lambda > 100\mu$m. Although this view is widespread, its validity has always been questioned. Sanders et al. (1989\nocite{Sanders89}) was amongst the first to do so, concluding that AGN must contribute to dust heating over kiloparsec (kpc) scales. More recently, Symeonidis et al. (2016\nocite{Symeonidis16}; hereafter S16) examined this idea in more detail, using a sample of nearby QSOs to derive an intrinsic average AGN SED from the optical to the submm. S16 showed that powerful AGN can be responsible for most of the FIR emission in the bolometric energy output of QSOs, implying that they can potentially heat dust at kpc scales. Based on these results, Symeonidis (2017\nocite{Symeonidis17}; hereafter S17) subsequently reported that the infrared spectral energy distributions (SEDs) of the most powerful QSOs are entirely AGN-dominated without the need of a star-formation component to explain any part of their energetic output, noting that the star-formation rates (SFRs) of these sources cannot be estimated reliably through broadband photometry alone. These results suggested that the FIR is not an unequivocal tracer of the SFR and that in sources with powerful AGN it is primarily indicative of AGN power.

\begin{table}
\centering
  \caption{Numbers of QSOs from each survey contributing to the
     sample in the $0.5<z<2.6$ range. Objects
    detected in FIRST are removed.}
\label{tab:QSOs}
\begin{tabular}{lrr} \hline
Survey & Number & 0.5$<z<$2.65 \\ \hline
DR10 & 3182 & 2249  \\
DR7 & 1327 & 1314 \\
2SLAQ & 1393 & 1384  \\
2QZ & 1371 & 1360  \\
KX & 60 & 53 \\
Total & 7333 & 6360  \\  
\hline
\end{tabular}
\end{table}

\begin{figure}
\epsfig{file=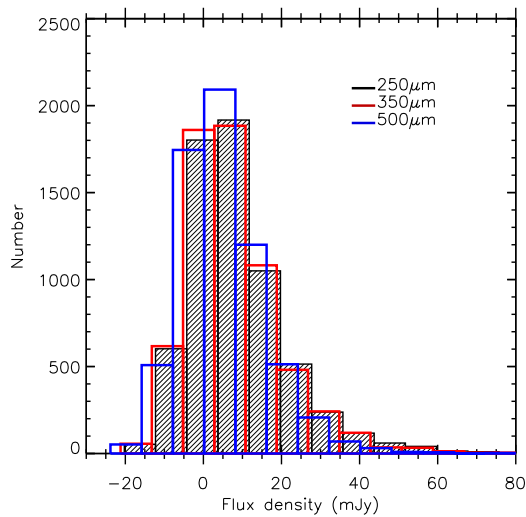,width=0.95\linewidth} 
\caption{The flux density distribution in the three SPIRE bands for the parent sample of QSOs; we remind the reader that we extract a SPIRE measurements for all QSOs (see section \ref{sec:sample})}
\label{fig:fluxdist}
\end{figure}

With the aim of investigating the putative relationship between star formation and
nuclear activity, we have assembled a large sample of optically-selected
QSOs at $z<3$, spanning more than two orders of magnitude in
accretion luminosity, with associated observations in the FIR. 
Our work differs from most other works in that we do not assume a-priori
that the FIR emission is predominantly powered by star-formation. 
The QSO sample and multi-wavelength data are described in 
Section \ref{sec:sample}. We describe how we determine the relative contributions to the FIR flux 
from the AGN and from the host in Section \ref{sec:method}, and the results are presented in
Section \ref{sec:results}. Discussion and conclusions are presented
in Sections \ref{sec:discussion} and \ref{sec:conclusions}.
Concordance cosmology with $H_{0} = 70$ km s$^{-1}$ Mpc$^{-1}$, $\Omega_{\rm M} = 0.3$,
$\Omega_{\Lambda} = 0.7$ is assumed.

\section{Sample selection }
\label{sec:sample}

The FIR photometry for the current work is from the
\textit{Herschel}\footnote{\textit{Herschel} is 
  an ESA space observatory with science instruments provided by
  European-led Principal Investigator consortia and with important
  participation from NASA} Space Observatory (Pilbratt et al. 2010\nocite{Pilbratt10}),
specifically the \textit{Herschel} Astrophysical Terahertz Large Area
Survey (H-ATLAS, Eales et al. 2010\nocite{Eales10a}), data release 1 (DR1).  DR1 (Valiante et al. 2016\nocite{Valiante16}) covers $\sim$162 deg$^2$ on three equatorial fields centred on RA = 9, 12, and 15h, which have extensive multiwavelength ancillary data and are also covered by the GAMA survey (Driver et al. 2009\nocite{Driver09}).

Within the area covered by the aforementioned H-ATLAS fields, we obtain QSOs from the following surveys: the Sloan Digital
Sky Survey (SDSS, York et al. 2000\nocite{York00}) Data Release 7 (DR7, Schneider et al. 2010\nocite{Schneider10}) and Data Release 10 (DR10,
P\^aris et al. 2014\nocite{Paris14}) catalogues, the 2dF SDSS LRG and QSO survey (2SLAQ,
Croom et al. 2009\nocite{Croom09}) and the 2dF QSO Redshift Survey (2QZ, Croom et al. 2004\nocite{Croom04}), all of which are by selection unobscured (i.e. broad lines or blue-bump; see the individual surveys for details). In addition to these large
surveys, we add unobscured QSOs selected in the NIR $K-$band by the KX
selection (KX QSOs, see Maddox et al. 2012\nocite{Maddox12} for a description of
the selection technique).

For objects common to more than
one survey, the order of preference in selecting them for our sample is SDSS DR10, SDSS DR7, 2SLAQ,
2dF, and KX. 
Note that the optical selection which results in a sample consisting solely of unobscured QSOs enables a direct and robust measure of the AGN power. Although by definition we are selecting against optically-obscured sources, this does not present a bias regarding the FIR: there is no evidence of a difference between the FIR-derived properties of type 1 and type 2 AGN (e.g. Rovilos et al. 2012; Suh et al. 2019\nocite{Suh19}, Zou et al. 2019\nocite{Zou19}) and as a result, we expect our results to be applicable to all AGN.

\begin{figure}
\epsfig{file=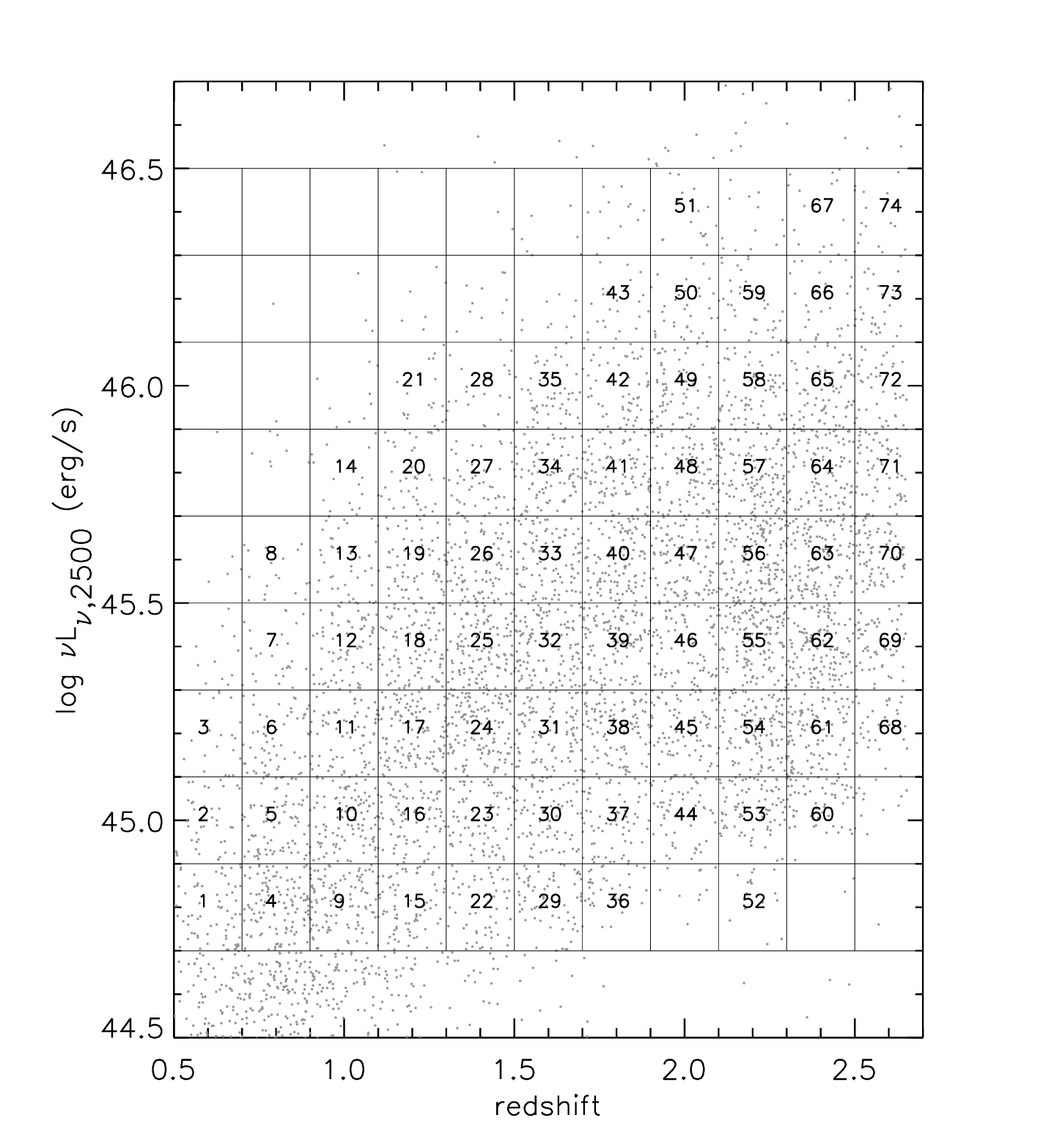,width=0.99\linewidth} 
\caption{Our sample of QSOs in log\,$\nu L_{\nu, 2500}$ - redshift space. The grid indicates our binning in 0.2\,dex in both $L$ and $z$, whereas the numbers correspond to the bin number (74 bins in total). Only bins with 15 sources or more are used in our analysis. There is a lower redshift cut of z=0.5 and a lower luminosity cut of log\,[$\nu L_{\nu, 2500}$\,(erg/s)]=44.7 in order to exclude sources where the host galaxy contributes to the optical/near-IR photometry. There is also an upper redshift cut of $z=2.65$ to ensure that rest-frame 2500\AA\,falls within the SDSS photometric bands. Our final sample consists of 5391 objects. }
\label{fig:bins}
\end{figure}

We extract the 250, 350 and 500$\mu$m SPIRE (Griffin et al. 2010\nocite{Griffin10}) flux densities at the pixel containing the optical position of \textit{each} QSO, using the mean-subtracted SPIRE maps processed by the \textit{Herschel} Extragalactic Legacy Project (HELP, Oliver et al. in prep; see also Shirley et al. 2019\nocite{Shirley19})\footnote{The HELP project focus is homogenising data from all \textit{Herschel} extragalactic surveys, covering over 1300\,deg$^2$ http://herschel.sussex.ac.uk; see also Vaccari 2016\nocite{Vaccari16}. Data available at  http://hedam.lam.fr/HELP/}. We double check that the mean of the maps is indeed zero by averaging random positions on the map. Note that this method of measuring flux at the central pixel reduces the effects of clustering on the measured flux of QSOs, compared to fitting a full PSF, with the flux boosting due to clustering effects estimated to be of the order of 8 per cent at 250$\mu$m (e.g. B{\'e}thermin et al. 2012\nocite{Bethermin12}). The distribution in flux density is shown in Fig. \ref{fig:fluxdist}. Across all 3 H-ATLAS fields we calculate the standard deviation of the distribution in pixel values to be 9.7, 9.2 and 9.2\,mJy at 250, 350 and 500$\mu$m respectively, corresponding to a QSO $3\sigma$ detection rate of 7, 7 and 3 per cent.

The combination of surveys results in a large dynamic range in luminosity and redshift, and potential biases regarding the amalgamation of the different surveys are discussed in section \ref{sec:check}. 
QSOs with a counterpart in the Faint Images of the Radio Sky at Twenty-centimeters survey (FIRST, Becker et al. 1995\nocite{Becker95}) are excluded, to avoid contamination of the FIR from synchrotron emission. We supplement the SDSS photometry ($ugriz$) with NIR $YJHK$ photometry from either the UKIRT Infrared Deep Sky Survey (UKIDSS) Large Area Survey (LAS,
Lawrence et al. 2007\nocite{Lawrence07}) or the Visible and Infrared Survey Telescope
for Astronomy (VISTA) Kilo-degree Infrared Galaxy Survey (VIKING),
using the Wide Field Camera (WFCAM) Science Archive (WSA,
Hambly et al. 2008\nocite{Hambly08}) and VISTA Science Archive (VSA, Cross et al. 2012\nocite{Cross12}) cross-matching the data within a 1\,arcsec radius. 
Furthermore, we cross-match our sample to the WISE (Wright et al. 2010\nocite{Wright10}) database within 1\,arcsec, retrieving photometry at 3.4, 4.6, 12, and 22$\mu$m.

We restrict the full sample of QSOs to objects in the $0.5\le z\le 2.65$ range so that rest-frame 2500\AA\,falls within the SDSS photometric bands. Choosing 2500\AA\,allows us to probe within the maximum of QSO emission while excluding only a small amount of sources from our final sample. For each object in that redshift range we subsequently calculate their \textit{K}-corrected rest-frame 2500\AA\, luminosity ($\nu L_{\nu, 2500}$) by linearly interpolating the SDSS bands. We also calculate the rest-frame \textit{K}-corrected 5100\AA\,luminosity ($\nu L_{\nu, 5100}$) for each source, by interpolating all available optical/near-IR data. For the sources for which 5100\AA\, (rest frame) falls outside the wavelength range covered by the available data we use the median SDSS QSO SED from Vanden Berk et al. (2001\nocite{vandenBerk01}) to convert from 2500\AA\, to 5100\AA. 
The number of objects from each survey
contributing our QSO sample is listed in
Table \ref{tab:QSOs}.

We partition the $L-z$ plane into bins of 0.2 in $z$ and 0.2 in log\,$\nu L_{\nu, 2500}$, as shown in Fig. \ref{fig:bins}. Only bins with 15 sources or more are used in our analysis. Finally we restrict our work to QSOs with log\,[$\nu L_{\nu, 2500}$\,(erg/s)]$>$44.7 in order to exclude sources where the host galaxy contributes to the optical/near-IR photometry, which we find to be the case when examining the optical/NIR SEDs of lower luminosity sources. Our final sample consists of 5391 sources.

\section{SED decomposition }
\label{sec:method}

For each $L-z$ bin (Fig. \ref{fig:bins}), we have the average $\nu L_{\nu, 2500}$ and $\nu L_{\nu, 5100}$ (see section \ref{sec:sample}), and the average 250, 350 and 500$\mu$m luminosity. The averages in the SPIRE bands are straight-forward, since, as stated in section \ref{sec:sample}, all sources have a measurement in all SPIRE bands. Individual luminosities are computed at the redshift of each source. We calculate the 68 per cent confidence intervals on the average SPIRE luminosities in each bin, by bootstrapping with 5000 iterations. 

If all sources in any given $L-z$ bin, are also detected in any of the 13 optical/NIR/MIR bands ($u,g,r,i,z$, $Y,J,H,K$, W1, W2, W3, W4), we subsequently compute the average luminosity in that band, in order to supplement our mean SED in that bin. This serves only as a consistency check when building the sources' SEDs (see Appendix A for the SEDs) and hence the incompleteness of data in these bands does not affect our results.

The AGN contribution to the average emission of sources in each bin is represented by the intrinsic AGN SED from S16. We acknowledge that there has been some controversy regarding the S16 SED (Lyu $\&$ Rieke 2017; Lani, Netzer $\&$ Lutz 2017\nocite{LNL17}; Stanley et al. 2018\nocite{Stanley18}; Schulze et al. 2019\nocite{Schulze19}; Xu et al. 2020\nocite{Xu20}) and thus refer the reader to Symeonidis (2022; hereafter S22\nocite{Symeonidis22}) where this is addressed. The S16 SED represents the average optical-submm broadband emission from AGN and was derived using a sample of optically luminous ($\nu L_{\nu 5100} > 10^{43.5}$ erg/s) unobscured, radio-quiet, $z<0.18$ QSOs from the Palomar Green (PG) survey. To obtain the intrinsic AGN emission, S16 subtracted the contribution from star-formation in the infrared, which was determined for each PG QSO as follows: the luminosity of the 11.3$\mu$m polycyclic aromatic hydrocarbon (PAH) feature in the QSO's MIR spectrum was matched to the SED template from the Dale $\&$ Helou (2002) library with the closest PAH luminosity (see Shi et al. 2007\nocite{Shi07}), a method which accurately reproduced the measured IR SED when tested on samples of star-forming galaxies (also see S22). 

As discussed in S22, the S16 AGN SED is an average so it is ideally used on averaged data and, here, we normalise it to the mean 5100\AA\,luminosity of each bin. Although the $\nu L_{\nu, 5100}$ range probed extends to higher luminosities and redshifts than that of the PG QSOs used to build the S16 SED, there is evidence that the intrinsic optical to FIR ratio of AGN is broadly independent of AGN power and redshift (see S16; S17; S22), consistent with the observation that the UV-to-mid-IR SEDs of QSOs also do not evolve as a function of redshift or AGN luminosity (e.g. Hao et al. 2014). We are thus confident that the S16 SED is a good measure of the intrinsic AGN emission for the entire range of AGN luminosities and redshifts probed in this work.

Once the AGN template is normalised to $\nu L_{\nu, 5100}$, the AGN FIR luminosity at $\lambda=250\mu m/(1+z)$ is subsequently subtracted from the average luminosity measured at that wavelength in each bin, in order to get the luminosity that can be attributed to star formation. To this we normalise a template from the Chary $\&$ Elbaz (2001\nocite{CE01}; hereafter CE01) star-forming SED library, by selecting the template with the closest luminosity in order to have the appropriate shape of far-IR emission and hence dust temperature. 
This approach simplifies the process of assigning a star-forming SED to each bin, when there are not enough data points for model fitting to be meaningful. ULIRG SED templates are excluded, as they were built on the SEDs of local ULIRGs which, as shown in Symeonidis $\&$ Page (2019\nocite{SP19}), have a significant AGN contribution. Since we use a local template library for all redshift bins, one concern might be the potential evolution in the average dust temperatures of galaxies (e.g. Symeonidis et al. 2013\nocite{Symeonidis13a}). However, Symeonidis et al. (2013) showed that evolution in dust temperatures between $z\sim0$ and $z\sim2$ is primarily confined to ULIRGs, hence by excluding them here from our selection of star-forming SED templates, we mitigate any issues that potentially arise because of this. Moreover, in section \ref{sec:check}, we show that our results are not sensitive to the choice of template. 

In each $L-z$ bin we now have the AGN SED component and the host galaxy SED component. The average IR luminosity from star-formation ($L_{\rm IR, SF}$) is computed by integrating the normalised CE01 template and the average AGN IR luminosity ($L_{\rm IR, AGN}$) is computed by integrating the normalised S16 template, all in the 8--1000$\mu$m wavelength range. The error on $L_{\rm IR, SF}$ is computed as follows: 
$\sigma_{\rm LIR,SF}=\frac{\sqrt (\sigma_{\rm L250, tot}^2 +\sigma_{\rm L250, AGN}^2)}{L_{\rm 250, SF}}L_{\rm IR, SF}$, where $\sigma_{\rm L250, tot}$ is the error on the 250$\mu$m luminosity of the given bin calculated by bootstrapping as indicated earlier, $\sigma_{\rm L250, AGN}$ is the error on the 250$\mu$m luminosity from the AGN computed from the S16 AGN SED uncertainty and $L_{\rm 250, SF}$ is the 250$\mu$m luminosity from star-formation, computed by subtracting the AGN luminosity from the total luminosity at that wavelength, as described above. 
Adding $L_{\rm IR, AGN}$ and $L_{\rm IR, SF}$ gives the average total IR luminosity ($L_{\rm IR, tot}$) of the given $L-z$ bin, with its associated uncertainty derived by combining $\sigma_{\rm LIR, SF}$ and $\sigma_{\rm LIR, AGN}$ in quadrature, the latter computed from the uncertainty on the S16 AGN SED.

\begin{figure*}
\begin{tabular}{c|c|c}
\epsfig{file=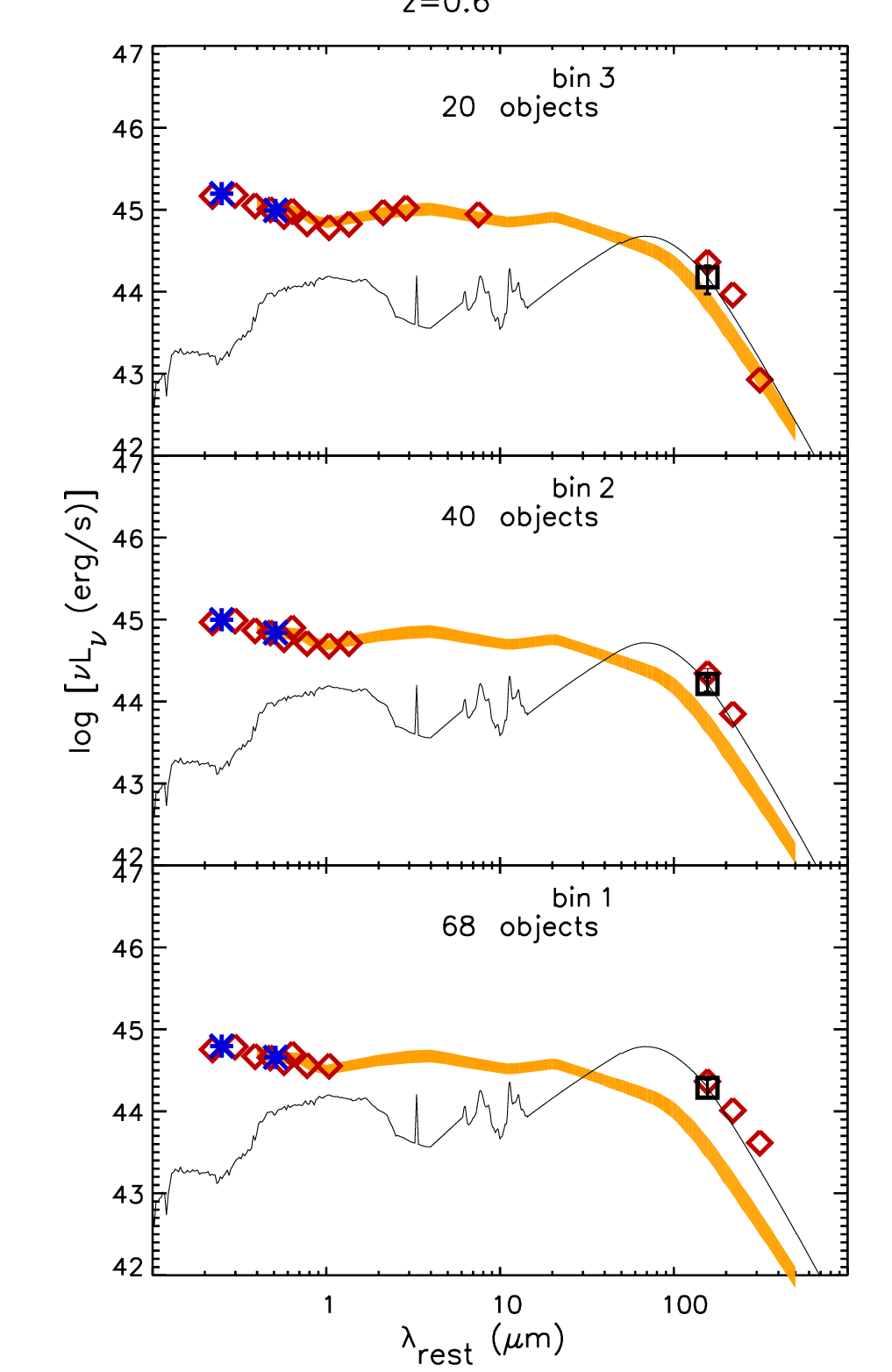,width=0.32\linewidth} & \epsfig{file=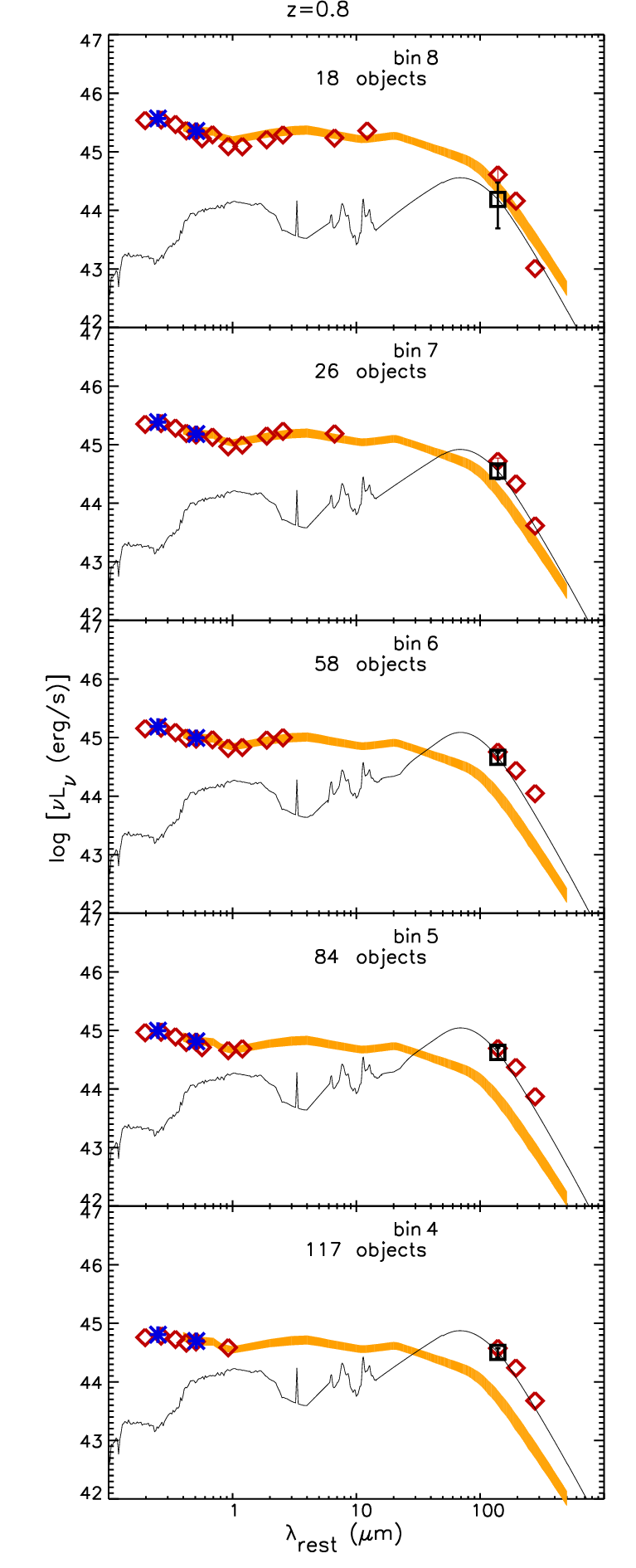,width=0.32\linewidth} & \epsfig{file=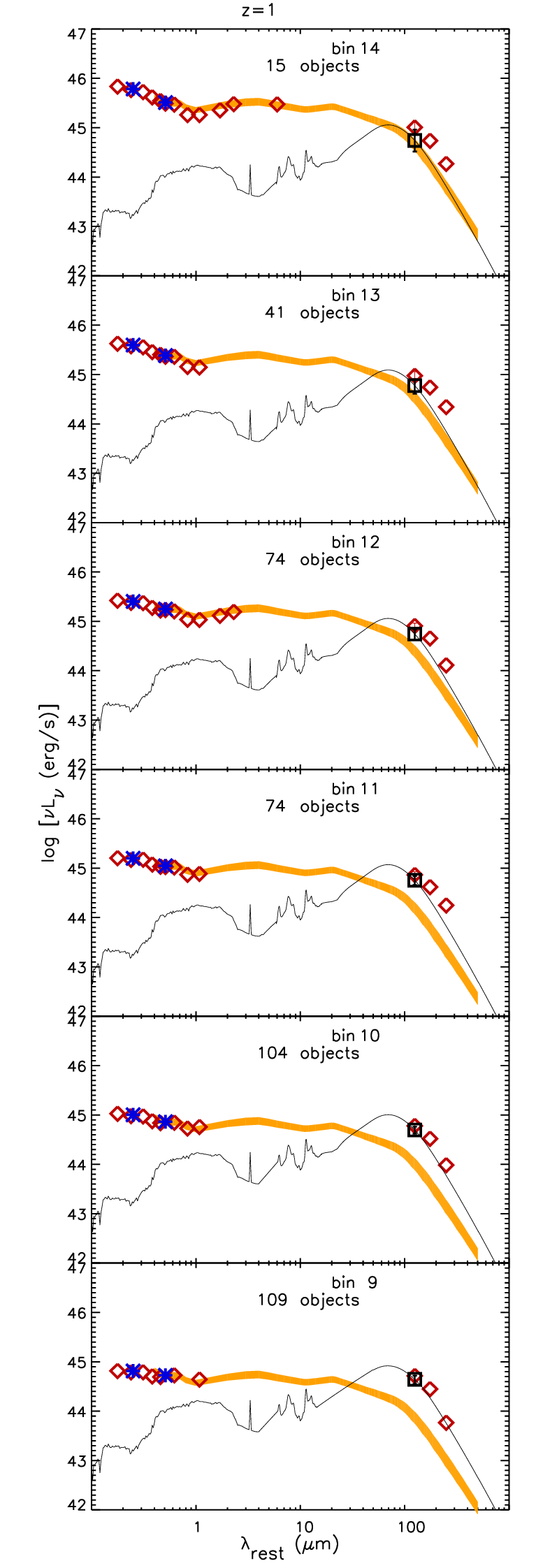,width=0.32\linewidth} \\ 
\end{tabular}
\caption{SED fitting for bins at redshift 0.6, 0.8 and 1 (see Fig. \ref{fig:bins}). Bin number increases from bottom to top (so increasing luminosity from bottom to top) and redshift increases from left to right. The bin number is indicated on the top right hand corner of each panel. The red diamonds are the average luminosities in each band. The blue asterisks are the rest-frame $K$-corrected 2500\AA\, and 5100\AA\, luminosities. Normalised onto the latter is the S16 intrinsic AGN SED (orange curve), with the thickness indicating the 68 per cent confidence intervals. The open black squares are the AGN-subtracted 250$\mu$m luminosity on which the CE01 host galaxy SED template is normalised (black curve).}
\label{fig:aseds1}
\end{figure*}

\begin{figure*}
\begin{tabular}{c|c|c}
\epsfig{file=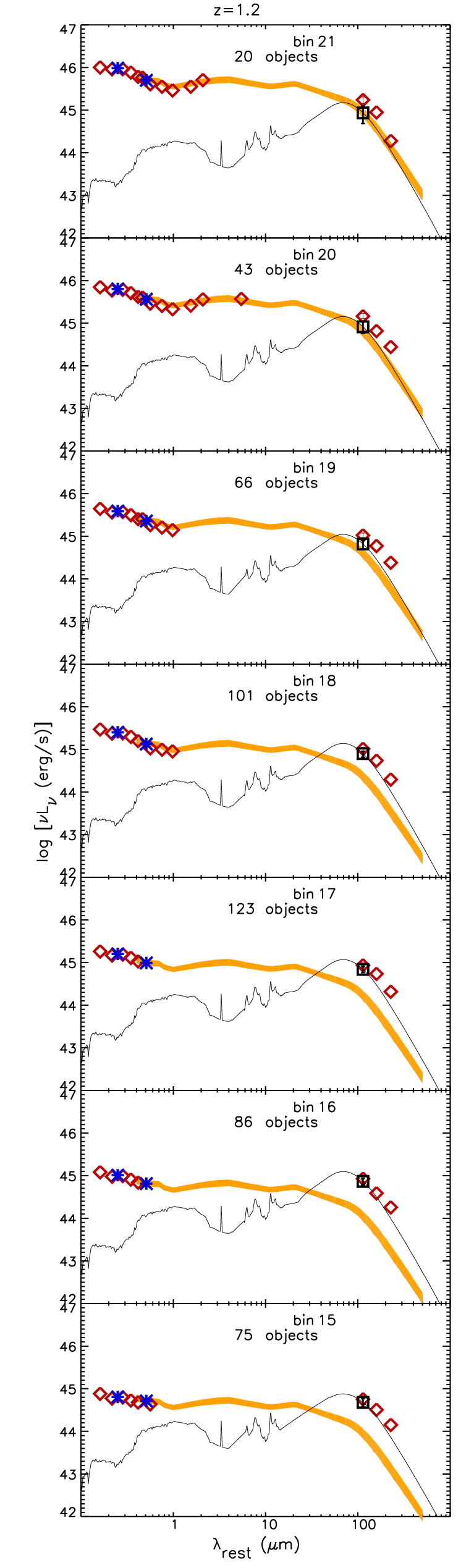,width=0.32\linewidth} & \epsfig{file=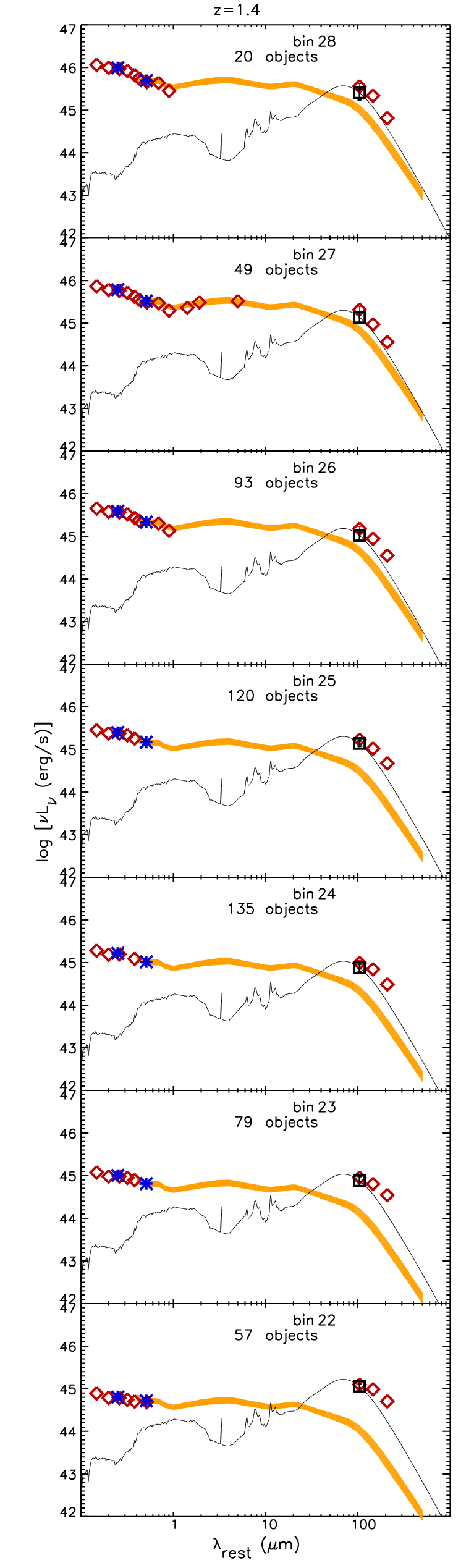,width=0.32\linewidth} & \epsfig{file=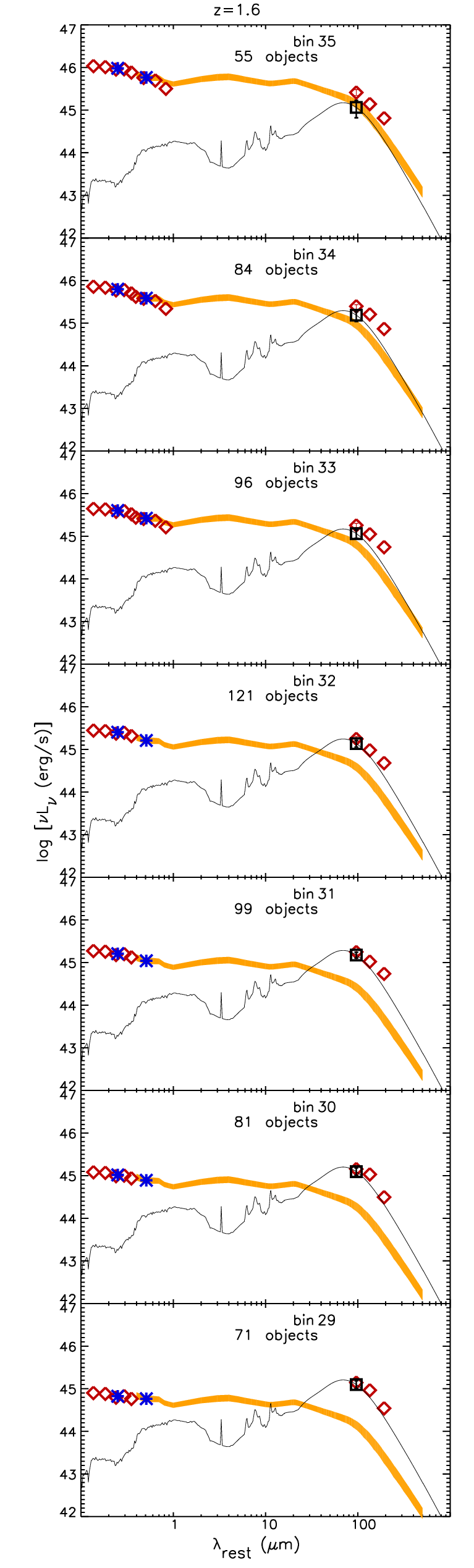,width=0.32\linewidth} \\ 
\end{tabular}
\caption{SED fitting for bins at redshift 1.2, 1.4 and 1.6 (see Fig. \ref{fig:bins}). Bin number increases from bottom to top (so increasing luminosity from bottom to top) and redshift increases from left to right. The bin number is indicated on the top right hand corner of each panel. The red diamonds are the average luminosities in each band. The blue asterisks are the rest-frame $K$-corrected 2500\AA\, and 5100\AA\, luminosities. Normalised onto the latter is the S16 intrinsic AGN SED (orange curve), with the thickness indicating the 68 per cent confidence intervals. The open black squares are the AGN-subtracted 250$\mu$m luminosity on which the CE01 host galaxy SED template is normalised (black curve).}
\label{fig:aseds2}
\end{figure*}

\begin{figure*}
\begin{tabular}{c|c|c}
\epsfig{file=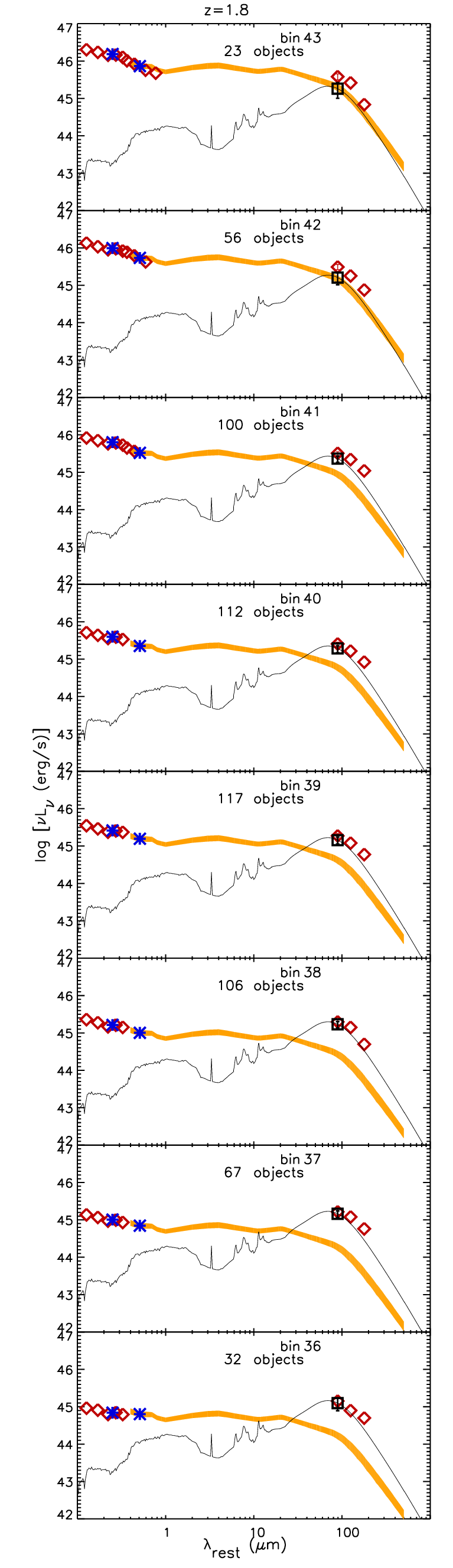,width=0.32\linewidth} & \epsfig{file=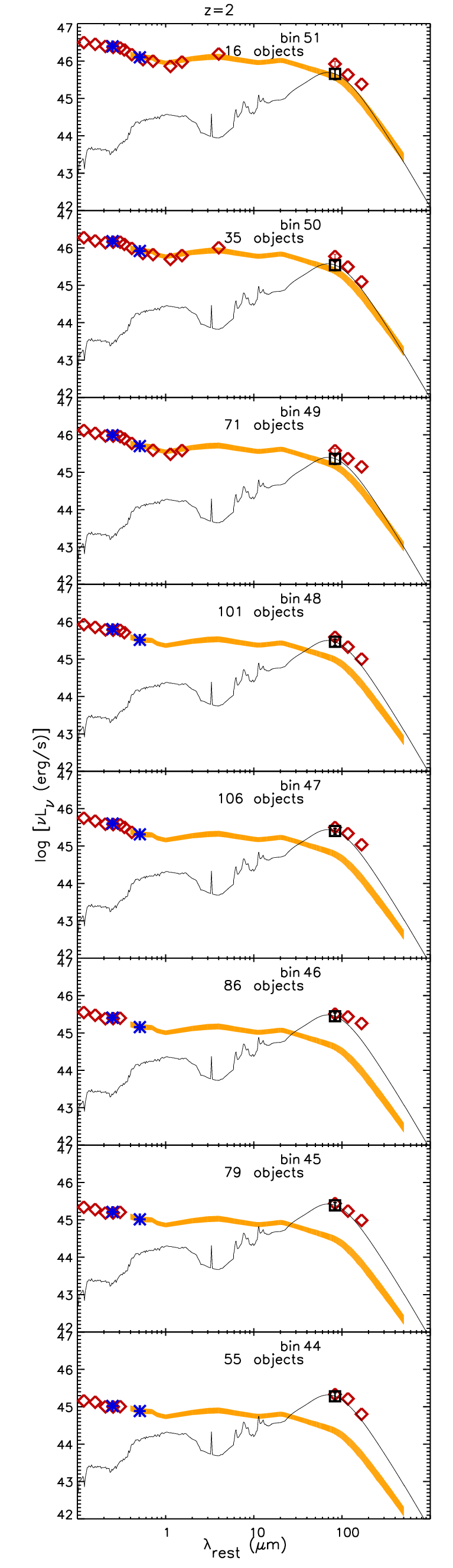,width=0.32\linewidth} & \epsfig{file=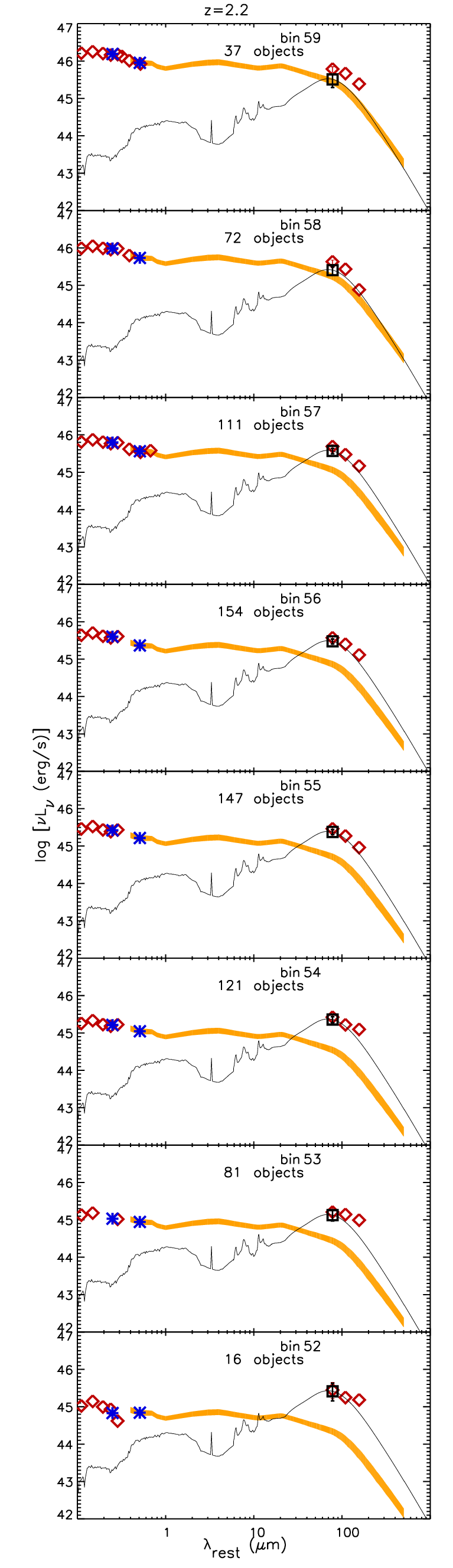,width=0.32\linewidth} \\ 
\end{tabular}
\caption{SED fitting for bins at redshift 1.8, 2 and 2.2 (see Fig. \ref{fig:bins}). Bin number increases from bottom to top (so increasing luminosity from bottom to top) and redshift increases from left to right. The bin number is indicated on the top right hand corner of each panel. The red diamonds are the average luminosities in each band. The blue asterisks are the rest-frame $K$-corrected 2500\AA\, and 5100\AA\, luminosities. Normalised onto the latter is the S16 intrinsic AGN SED (orange curve), with the thickness indicating the 68 per cent confidence intervals. The open black squares are the AGN-subtracted 250$\mu$m luminosity on which the CE01 host galaxy SED template is normalised (black curve).}
\label{fig:aseds3}
\end{figure*}

\begin{figure*}
\begin{tabular}{c|c|c}
\epsfig{file=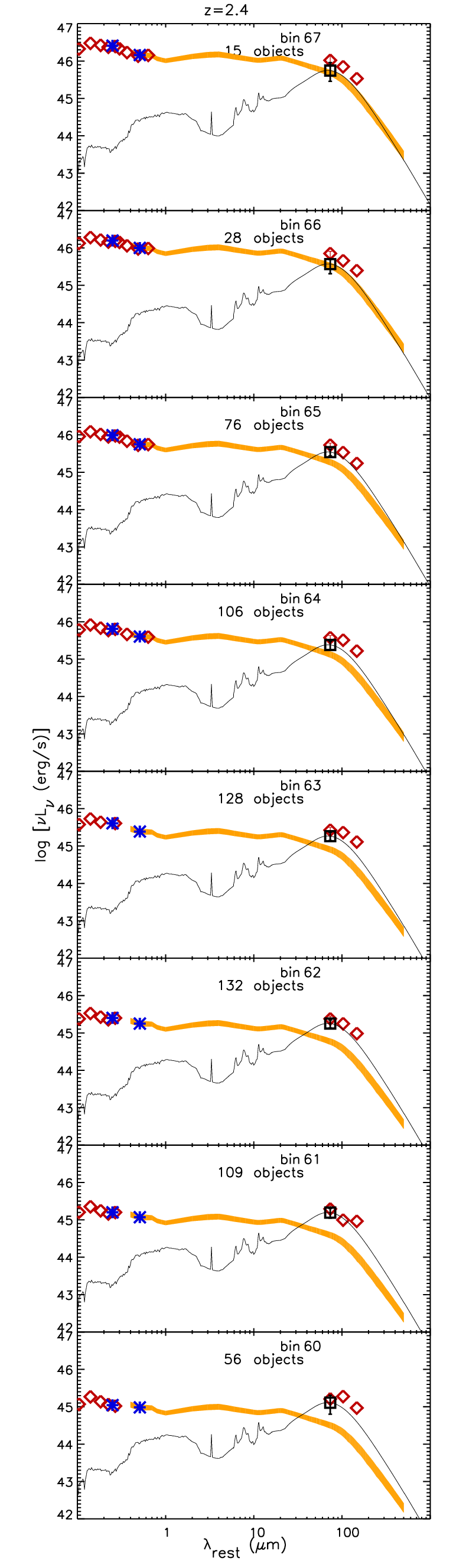,width=0.32\linewidth} & \epsfig{file=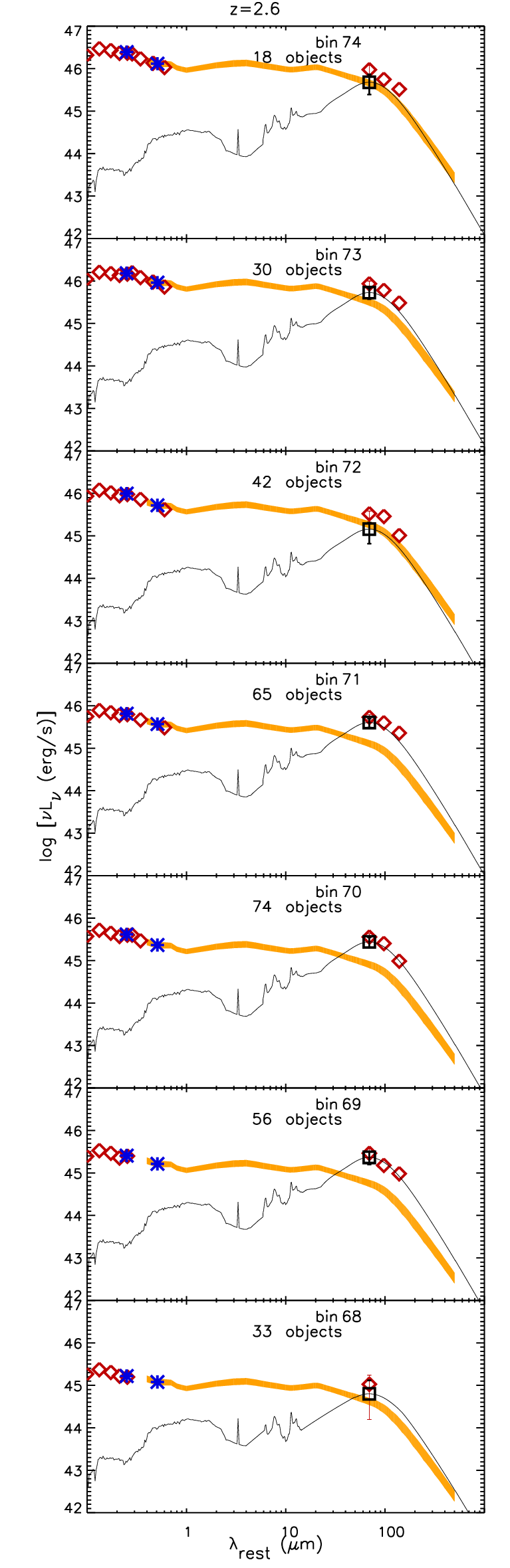,width=0.32\linewidth} & \\ 
\end{tabular}
\caption{SED fitting for bins at redshift 2.4 and 2.6 (see Fig. \ref{fig:bins}). Bin number increases from bottom to top (so increasing luminosity from bottom to top) and redshift increases from left to right. The bin number is indicated on the top right hand corner of each panel. The red diamonds are the average luminosities in each band. The blue asterisks are the rest-frame $K$-corrected 2500\AA\, and 5100\AA\, luminosities. Normalised onto the latter is the S16 intrinsic AGN SED (orange curve), with the thickness indicating the 68 per cent confidence intervals. The open black squares are the AGN-subtracted 250$\mu$m luminosity on which the CE01 host galaxy SED template is normalised (black curve).}
\label{fig:aseds4}
\end{figure*}

\begin{figure}
\epsfig{file=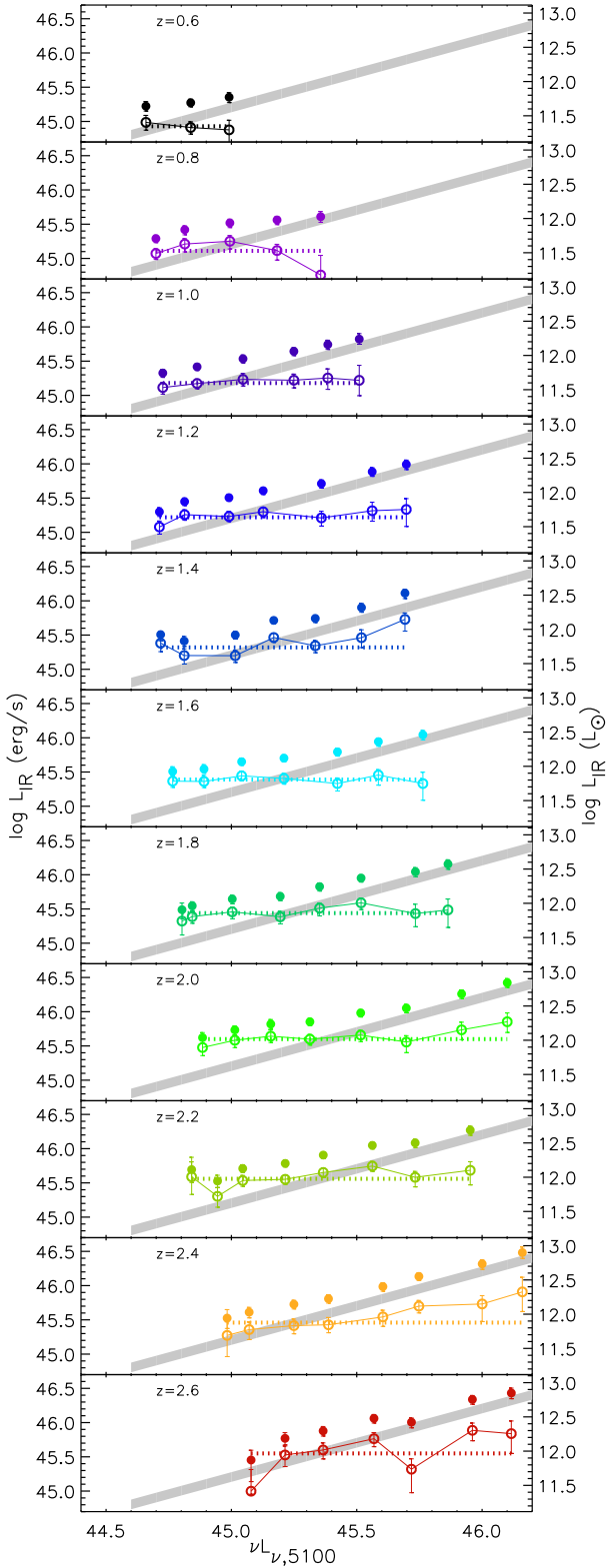,width=0.95\linewidth} 
\caption{$L_{\rm IR}$ versus $\nu L_{\nu, 5100}$ for each redshift bin. The central redshift of the bin is indicated on the top left corner of each panel. The shaded diagonal line represents the AGN locus as defined by the S16 intrinsic AGN SED. The width of the shading represents the 1$\sigma$ boundaries. The filled circles represent the average $L_{\rm IR, tot}$ in each $L-z$ bin, whereas the empty circles are $L_{\rm IR, SF}$. The dotted line is the best fit constant for the values of $L_{\rm IR, SF}$ (see table \ref{tab:sfr}).}
\label{fig:AGNSFR_separate}
\end{figure}

\begin{figure}
\epsfig{file=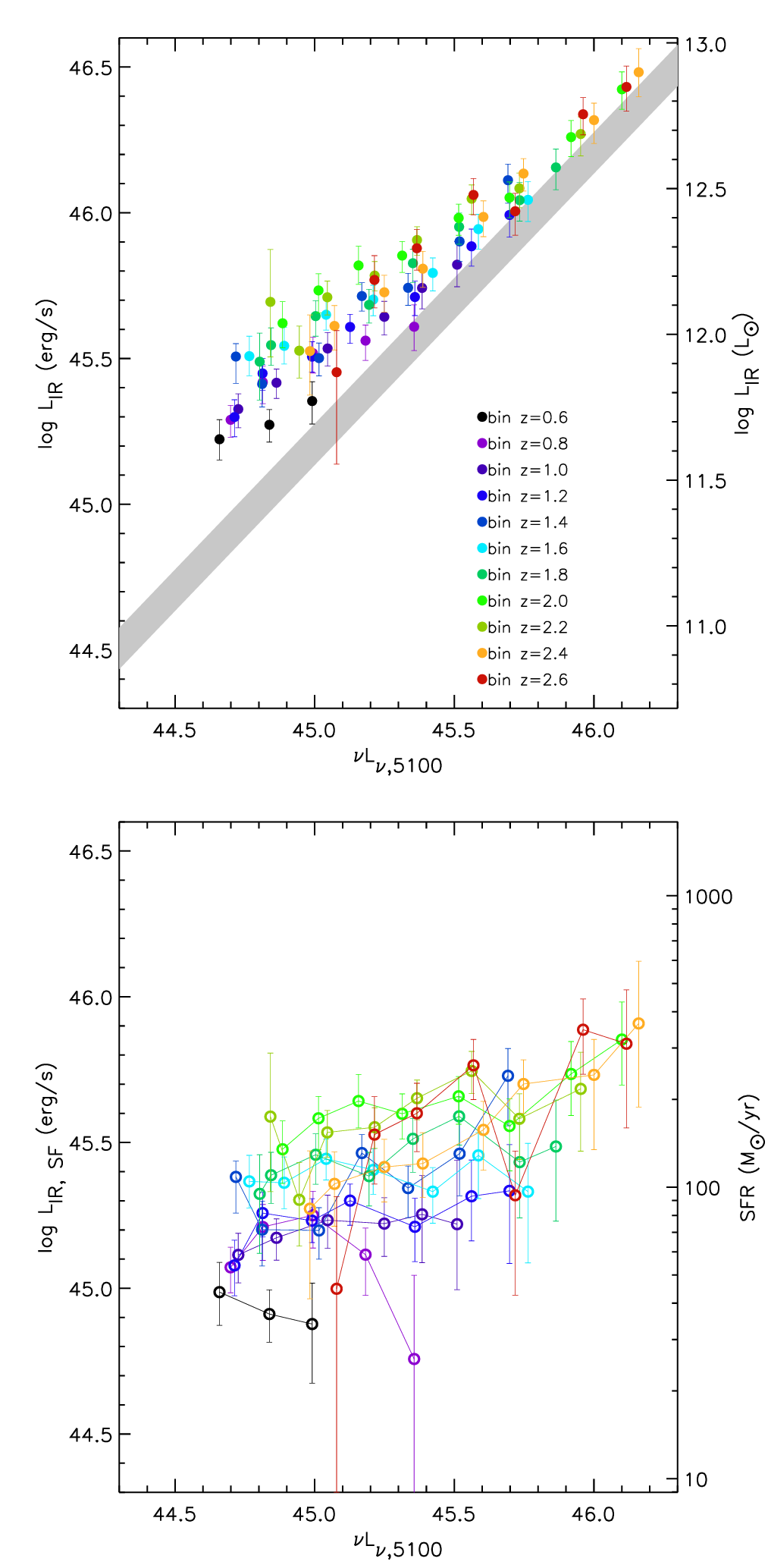,width=0.99\linewidth} 
\caption{Top panel: $L_{\rm IR}$ versus $\nu L_{\nu, 5100}$ with all redshift bins shown together. Lower panel: $L_{\rm IR, SF}$ versus $\nu L_{\nu, 5100}$. Equivalent star-formation rates are shown on the right y-axis derived using the Kennicutt (1998) calibration. The shaded diagonal line represents the AGN locus as defined by the S16 intrinsic AGN SED (top panel); the width of the shading represents the 1$\sigma$ boundaries.}
\label{fig:AGNSFR}
\end{figure}

\section{Results}
\label{sec:results}

The average SEDs of each bin (Fig. \ref{fig:bins}) are shown in Figs \ref{fig:aseds1}, \ref{fig:aseds2}, \ref{fig:aseds3}, \ref{fig:aseds4}. The S16 SED template which is normalised at 5100\AA\, (see section \ref{sec:method}) is seen to agree well with the average optical and MIR colours (where available) for the binned SEDs. It is interesting to note that the `FIR bump' ($>80\mu m$) in each binned SED is being filled progressively more by the AGN IR emission as the bin luminosity increases, and eventually the average SED settles onto the S16 intrinsic AGN SED. This effect is quantitatively shown in Fig. \ref{fig:AGNSFR_separate}, where $L_{\rm IR, tot}$ is plotted against $\nu L_{\nu, 5100}$. The redshift bins are plotted separately for clarity in Fig. \ref{fig:AGNSFR_separate}, but also all together in Fig. \ref{fig:AGNSFR} (top panel). At all redshifts the same trend is observed: at low AGN power, $L_{\rm IR, tot}$ shows a larger offset from the AGN locus than at high AGN power and eventually $L_{\rm IR, tot}$ converges onto the AGN locus. Correcting $L_{\rm IR, tot}$ for the AGN contribution in order to get $L_{\rm IR, SF}$ (see section \ref{sec:method}), shows a different trend entirely: $L_{\rm IR, SF}$ is independent of AGN power (see Fig. \ref{fig:AGNSFR_separate} for the separate redshift bins and Fig. \ref{fig:AGNSFR}, lower panel, for all bins together). At all redshifts, we examine whether $L_{\rm IR, SF}$ can be assumed constant over the range of $\nu L_{\nu, 5100}$ spanned here, by $\chi^2$ fitting a constant to the data (see Fig \ref{fig:AGNSFR_separate}). We find that a single value of $L_{\rm IR, SF}$ is an acceptable fit to the data in each redshift range at the 95 per cent confidence level. 

\begin{figure}
\epsfig{file=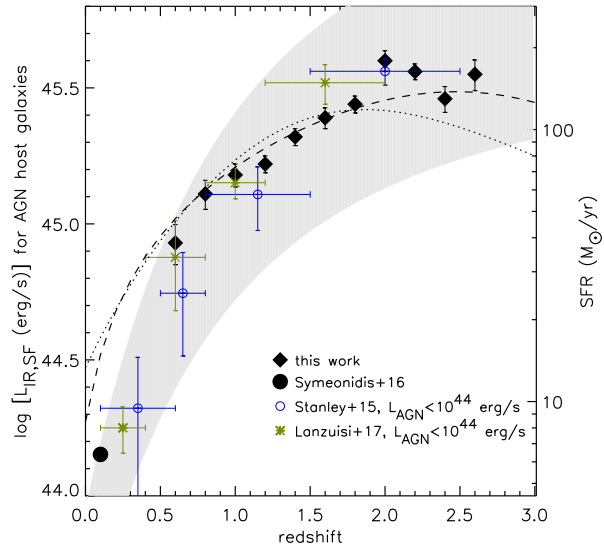,width=0.99\linewidth} 
\caption{The average $L_{\rm IR, SF}$ (left y-axis) or SFR (right y-axis; derived using the Kennicutt 1998 calibration) of QSOs in each of our redshift bins (black diamonds). Each point is  the constant fitted on our data in Fig \ref{fig:AGNSFR_separate}. The black circle is the average $L_{\rm IR, SF}$ for the sample of PG QSOs of S16. Blue circles and green asterisks represent the average $L_{\rm IR, SF}$ we compute from the Stanley et al. (2015) and Lanzuisi et al. (2017) results respectively --- in both cases the horizontal error bars indicate the size of the redshift bin. For both the Stanley et al. and the Lanzuisi et al. samples, we only use their log\,[$L_{\rm X, AGN}(\rm erg/s)]<$44 sources, since at low AGN luminosities the AGN does not contribute substantially to the IR. The dotted line represents the star-formation rate density (SFRD) as a function of redshift from Madau $\&$ Dickinson (2014) assuming a Salpeter IMF and converted to luminosity units using the Kennicutt (1998) calibration. The dashed line is the Beacom $\&$ Hopkins (2006) SFRD again assuming a Salpeter IMF and converted to luminosity units using the Kennicutt (1998) calibration. Note that these relations are not fitted to our data, but scaled by an arbitrary amount for the purpose of comparing their shape with the trend seen in our results. Finally the grey shaded region represents the $SFR-M_{\star}$ `main sequence' relation from Speagle et al. (2014) in the log\,[M$_{\star}$ (M$_{\odot}$)]=10.5 (lower boundary) to the log\,[M$_{\star}$ (M$_{\odot}$)]=11.5 (upper boundary) range. }
\label{fig:sfr}
\end{figure}

In Fig. \ref{fig:sfr} we plot the average $L_{\rm IR, SF}$ determined for each redshift bin, i.e. the constants fitted on our data in Fig \ref{fig:AGNSFR_separate}. We also compare with the results from Stanley et al. (2015\nocite{Stanley15}) and Lanzuisi et al. (2017\nocite{Lanzuisi17}), converting the AGN power from X-rays to the rest-frame 5100\AA, using the relation from Maiolino et al. (2007\nocite{Maiolino07}). From these samples we only chose the low luminosity AGN, i.e. $L_{\rm X, AGN}<10^{44}$\,erg/s, where the contribution of the AGN to the IR is minimal (see discussion). We also plot the average $L_{\rm IR, SF}$ computed in S16 with a sample of nearby lower luminosity PG QSOs. It is evident that $L_{\rm IR, SF}$ increases as a function of redshift, plateauing at about $2<z<3$. The increase from $z\sim0$ to $z\sim3$ is about a factor of 30, the equivalent increase in SFR from $\sim$6 to $\sim$200\,M$_{\odot}$/yr, using the Kennicutt (1998\nocite{Kennicutt98}) calibration. As shown, the redshift evolution in average SFR of AGN hosts is similar to the increase in star-formation rate density (SFRD) as a function of redshift (Hopkins $\&$ Beacom 2006\nocite{HB06}; Madau $\&$ Dickinson 2014\nocite{MD14})--- note that we have scaled the SFRD by an arbitrary amount for the purpose of comparing its shape with the trend seen in our results.  The increase in $L_{\rm IR, SF}$ with redshift for AGN hosts is also consistent with the redshift evolution of the SFR--$M_{\star}$ relation (`main sequence of star-formation') in the log\,[$M_{\star}$ (M$_{\odot}$)]=10.5-11.5 range, taken from Speagle et al. (2014\nocite{Speagle14}). These findings indicate that AGN host galaxies are indistinguishable from the general galaxy population in terms of their average SFR and its evolution with redshift. 

\subsection{Consistency checks}
\label{sec:check}
We repeat our analysis separately for each of the 5 QSO surveys that we combine (see section \ref{sec:sample}) in order to examine potential selection biases introduced by the different surveys. We find no evidence for bias --- sources from the different surveys display similar optical/NIR QSO properties and the results show the same trends, hence we are confident that our results are not affected by the selection criteria employed in the QSO samples that we have used. 

To understand the effect of our choice of star-forming template on the final $L_{\rm IR}$ estimates, we re-calculate $L_{\rm IR}$, $L_{\rm IR, SF}$, $L_{\rm IR, AGN}$ by choosing arbitrary templates from the CE01 library, rather than matching them in luminosity as described in section \ref{sec:method}. We find that our computed values change only marginally and within the original computed statistical uncertainties.

\begin{table}
\centering
  \caption{Data for Fig \ref{fig:sfr} --- $L_{\rm IR, SF}$ and SFRs of QSOs. $L_{\rm IR, SF}$ is the constant fitted on our data in Fig \ref{fig:AGNSFR_separate}. The 1$\sigma$ error is computed according to the prescribed $\chi^2$ confidence intervals for one interesting parameter, namely $\chi_{\rm min}^2 + 1$. SFRs are computed using the Kennicutt 1998 calibration.}
\label{tab:sfr}
\begin{tabular}{lccc} \hline
$z$ & $L_{\rm IR, SF}$  & 1$\sigma$ error & SFR \\
 & ($\times 10^{45}$ erg/s) &($\times 10^{45}$ erg/s) & (M$_{\odot}$/yr)\\
\hline
           0.6&      0.85&      0.14  &     38.30\\
     0.8&       1.23&      0.16    &   57.97\\
      1&       1.51&      0.15     &  68.11\\
      1.2&       1.66&      0.12   &    74.68\\
      1.4&       2.09&      0.15   &    94.02\\
      1.6&       2.45&      0.22   &    110.46\\
      1.8&       2.75&      0.20   &    123.94\\
      2&       3.98&      0.35    &   179.15\\
      2.2&       3.63&      0.24      & 163.39\\
      2.4&       2.88&      0.31    &   129.78\\
      2.6&       3.55&      0.46    &   159.67\\
      \hline
      \end{tabular}
      \label{tab:sfr}
\end{table}

\section{Discussion}
\label{sec:discussion}

We have examined the SEDs of a sample of 5391 optically selected QSOs at $0.5<z<2.65$ with log [$\nu L_{\nu, 5100}$ (erg/s)]=44.7-46.5, binned in 74 luminosity-redshift bins. In each redshift range, we investigate the relation between the average $L_{\rm IR, tot}$ and the AGN power ($L_{\rm AGN}$) parametrised by $\nu L_{\nu, 5100}$. We note a correlation between $L_{\rm IR, tot}$ and $\nu L_{\nu, 5100}$ in all redshift bins. However, after separating the total IR emission into a host ($L_{\rm IR, SF}$) and an AGN ($L_{\rm IR, AGN}$) component, we find that $L_{\rm IR, SF}$ is consistent with being constant over the range of AGN power probed. Moreover we notice that the contribution of the AGN to the IR increases as a function of AGN luminosity, manifesting as a change in the average SED of each $L-z$ bin, from displaying a prominent far-IR bump, to eventually settling onto the S16 intrinsic AGN SED (see SEDs in Appendix A). 

We calculate the average SFR of optically unobscured QSOs and find that it increases from about 30\,M$_{\odot}$/yr at $z\sim0.5$ to a plateau of bit less than 200\,M$_{\odot}$/yr at $z\sim2.6$, broadly consistent with the increase in star-formation rate density between those epochs (Fig \ref{fig:sfr}). Our computed SFRs are consistent with the SFRs estimated from the X-ray selected AGN sample of Stanley et al. (2015) and Lanzuisi et al. (2017) of low AGN power ($L_{\rm X, AGN}<10^{44}$\,erg/s) where the AGN does not contribute substantially to the far-IR. Our results show that even at the peak of SFR density and AGN accretion rate density ($z\sim2$; e.g. Boyle $\&$ Terlevich 1998\nocite{BT98}), the average SFR of AGN host galaxies does not exceed 200\,M$_{\odot}$/yr. Moreover, the increase in $L_{\rm IR,SF}$ of AGN host galaxies with redshift is also consistent with what is expected from the increase in the normalisation of the $SFR-M_{\star}$ relation with cosmic time, in the $10^{10.5}-10^{11.5}$ stellar mass range. Note that the conversion from $L_{\rm IR, SF}$ to SFR is uncertain as it depends on assumptions regarding the initial mass function (IMF). A top heavy IMF, which is shown to be the case in certain environments (e.g. Romano et al. 2017\nocite{Romano17}; Zhang et al. 2018\nocite{Zhang18}), would significantly alter SFR estimates. Nevertheless, our findings regarding the trend in SFR with redshift, as well as the comparison between the average SFRs of different samples are not affected as we use a consistent $L_{\rm IR, SF}$-SFR calibration. 

Given our findings that the most luminous AGN have similar SFRs to their lower luminosity counterparts, it is possible that the high reported SFRs in some QSO samples (e.g. Pitchford et al. 2016\nocite{Pitchford16}; Harris et al. 2016\nocite{Harris16}; Netzer et al. 2016\nocite{Netzer16}; Duras et al. 2017\nocite{Duras17}) are a consequence of those works underestimating the AGN contribution in the far-IR, due to their choice of AGN SED (e.g. see S22).

\subsection{The relation between AGN power and IR emission}
Many works have investigated the putative correlation between AGN power and total IR luminosity ($L_{\rm AGN}$-$L_{\rm IR}$) or star-forming IR luminosity ($L_{\rm AGN}$-$L_{\rm IR, SF}$), while also looking into fundamental galaxy and AGN relations with respect to black hole mass, AGN variability and galaxy stellar mass (e.g. Shao et al. 2010\nocite{Shao10}; Bonfield et al. 2011; Kalfountzou et al. 2012; Rovilos et al. 2012\nocite{Rovilos12}; Rosario et al. 2012, 2013; Chen et al. 2013\nocite{Chen13}; Hickox et al. 2014\nocite{Hickox14}; Azadi et al. 2015\nocite{Azadi15}; G{\"u}rkan et al. 2015\nocite{Gurkan15}; Stanley et al. 2015\nocite{Stanley15}; Harris et al. 2016\nocite{Harris16}; Pitchford et al. 2016\nocite{Pitchford16}; Stanley et al. 2017\nocite{Stanley17}; Lyu $\&$ Rieke 2017\nocite{LR17}; Lani et al. 2017\nocite{LNL17}; Bianchini et al. 2019\nocite{Bianchini19}). Although the range of AGN power probed in these studies varies, the results are essentially the same: at low AGN luminosity, $L_{\rm AGN}$-$L_{\rm IR}$ is flat, whereas at high AGN luminosity, $L_{\rm AGN}$-$L_{\rm IR}$ steeply rises. Our findings agree with these studies regarding the shape of the $L_{\rm AGN}$-$L_{\rm IR}$ trend. Where we differ however, is the interpretation of this trend, which is sensitive to the amount of far-IR emission attributed to the AGN. If the contribution of AGN to the far-IR is either ignored or estimated using an AGN SED with the a-priori assumption that the AGN does not contribute significantly to the far-IR (e.g. Shao et al. 2010; Rosario et al. 2012; 2013; Stanley et al. 2015; 2017; see also S22), $L_{\rm AGN}$-$L_{\rm IR, SF}$ is largely the same as $L_{\rm AGN}$-$L_{\rm IR}$. The change in the $L_{\rm AGN}$-$L_{\rm IR, SF}$ relation from flat to rising then requires a, perhaps counter-intuitive, break in galaxy behaviour, in some cases interpreted as being due to a change in the mode of black hole and galaxy growth (e.g. secular versus merger-induced; Shao et al. 2010; Rosario et al. 2012; 2013) or coupled to the various underlying relations between SFR, stellar mass, black hole mass and Eddington ratio (e.g. Stanley et al. 2015; 2017). However, such interpretations do not offer an explanation for why there should be a break in galaxy behaviour at some arbitrary AGN luminosity. 
S16 had argued that, instead, the $L_{\rm AGN}$-$L_{\rm IR}$ relation could be explained simply as an increase in the AGN contribution to the infrared. Below we explicitly show this.

\begin{figure}
\epsfig{file=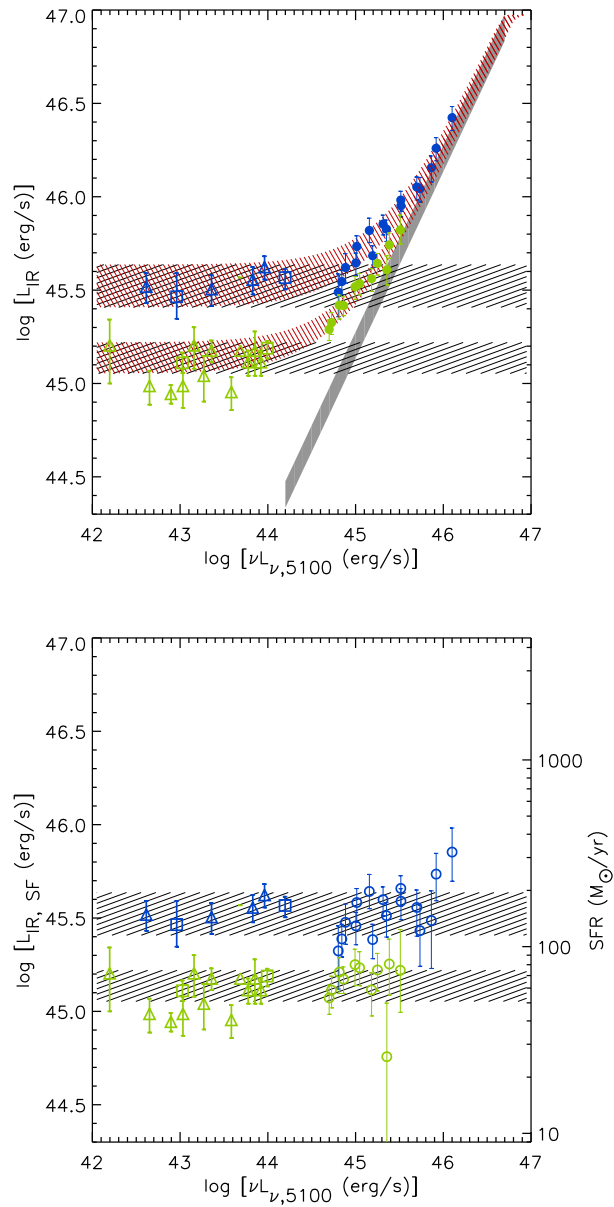,width=0.99\linewidth} 
\caption{\textit{Top panel}: $L_{\rm IR}$ vs AGN power at 5100\AA, combining our results (filled circles) with those of Stanley et al. (2015; open triangles) and Lanzuisi et al. (2017; open squares) for low luminosity (log\,[$L_{\rm X, AGN} (\rm erg/s)<$44]) AGN, in the same redshift ranges; green for $0.7<z<1.1$ and blue for $1.7<z<2.1$.
The horizontal black hatched regions represent the average $L_{\rm IR, SF}$ from our data at the quoted redshift ranges (see Fig \ref{fig:sfr}). The diagonal solid grey region represents the $L_{\rm IR, AGN}-\nu L_{\nu, 5100}$ correlation as defined by the S16 AGN SED. The width of the shading represents the 1$\sigma$ boundaries. Summing the horizontal and diagonal lines in each redshift range, produces the red hatched regions which trace the trend seen in the data over the entire range in AGN luminosity. \textit{Bottom panel}: Similar to the top panel, apart from, now, the infrared luminosity for our sample is corrected for the AGN contribution and hence $L_{\rm IR, SF}$ is plotted (open circles). Our data now matches the range in $L_{\rm IR, SF}$ seen in the Stanley et al. (2015) and Lanzuisi et al. (2017) samples of lower luminosity AGN. }
\label{fig:schem}
\end{figure}

Using the $0.7<z<1.1$ and $1.7<z<2.1$ redshift ranges to exemplify this point, Fig. \ref{fig:schem} (top panel) shows that the $L_{\rm AGN}$-$L_{\rm IR}$ relation can be reproduced by a superposition of two independent components: (A) the infrared emission from star-formation, unrelated to AGN power and (B) the infrared emission from AGN correlated with the AGN power measured in the optical or X-rays. To outline component A we plot our average $L_{\rm IR, SF}$ in each of the two redshift ranges (black hatched area) and component B is derived from the S16 intrinsic AGN SED (solid grey area). Summing the two reproduces the trends observed in the data: in the low AGN power regime, the host galaxy dominates $L_{\rm IR, tot}$, hence it is proportional to the SFR and the flat slope of the trend essentially reflects the lack of a relation between SFR and AGN power. Indeed the data from Stanley et al. (2015) and Lanzuisi et al. (2017) for low luminosity AGN in the same redshift ranges as our sources (also shown in Fig \ref{fig:sfr}) are seen to lie on the flat part of the trend. In the high AGN luminosity regime, probed by our data, $L_{\rm IR, tot}$ now has a significant AGN contribution and, as a result, it stops tracing the SFR and starts tracing the AGN power. Although, due to low number statistics we cannot probe the average IR emission for log\,[$\nu L_{\nu, 5100}$\,erg/s]$>$46.5 AGN, Fig. \ref{fig:schem} indicates that eventually the AGN makes up the entire $L_{\rm IR, tot}$, rendering the host galaxy a minor component in the infrared. Indeed S17 who studied the SEDs of the most luminous QSOs at $2<z<3.5$, found that only an upper limit to the SFR could be computed in such cases. Our results are also in agreement with Symeonidis $\&$ Page (2021; hereafter SP21\nocite{SP21}) who investigated the behaviour of the galaxy and AGN luminosity functions and concluded that there is a sharp drop in the reliability of IR-derived SFRs above a certain AGN power. Once the AGN contribution is subtracted (Fig. \ref{fig:schem}, bottom panel), our sample now overlaps the hatched regions, i.e. $L_{\rm IR, SF}$ for our sample of luminous AGN is consistent with the $L_{\rm IR, SF}$ of lower luminosity AGN. This suggests that there is no relation between $L_{\rm IR, SF}$ (or SFR) and AGN power --- $L_{\rm IR, SF}$ in AGN host galaxies is a function of redshift only (as also shown in Fig \ref{fig:sfr}). Moreover the agreement between our computed SFRs for unobscured AGN and those for X-ray selected AGN, some of which are obscured suggests that our conclusions are applicable to the AGN population in general. 

The final point we wish to make is one that relates to the potential suppression of star-formation by AGN. Various authors claim that evidence against powerful AGN suppressing star-formation stems from the fact that the estimated SFRs of such sources are high (e.g. Harrison et al. 2012\nocite{Harrison12}; Harris et al. 2016\nocite{Harris16}; Schulze et al. 2019\nocite{Schulze19}). However, earlier work by Page et al. (2012\nocite{Page12}) revealed a dearth of far-IR detections amongst the most luminous AGN at $z\sim1-3$, in line with our findings that the most luminous sources display the smallest far-IR bump, i.e. they have the shape of the AGN SED in the far-IR/sub-mm rather than the SED of a typical star-forming galaxy. It all connects to the idea that the most powerful AGN completely drown their host in the infrared and only an upper limit to the SFR can be calculated from IR broadband emission in such cases. Indeed, in the effort to verify the accuracy of IR-derived SFRs in AGN host galaxies, other SFR indicators must be explored, such as high resolution radio data and spectroscopy. Regarding the latter, the \textit{JWST} will offer the opportunity to examine polycyclic aromatic hydrocarbons (PAHs; Peeters et al. 2004\nocite{Peeters04}; Risaliti et al. 2006\nocite{Risaliti06}; Kennicutt et al. 2009\nocite{Kennicutt09}) in galaxies' mid-IR spectra as tracers of star formation in AGN host galaxies. Although our results  --- that SFRs in AGN host galaxies appear constant at any given redshift, irrespective of AGN luminosity --- seem to be at odds with a scenario whereby the most powerful AGN quench star-formation, we are missing a key galaxy property: the stellar mass. This is difficult to accurately compute for optically unobscured QSOs, but indeed, if the most luminous AGN are found in more massive galaxies than their low luminosity counterparts, the \textit{specific} SFR will show a declining trend with increasing AGN luminosity, evidence that star-formation is suppressed (e.g. Dubois et al. 2016\nocite{Dubois16}).

\section{Conclusions}
\label{sec:conclusions}

We have examined the SEDs of a sample of 5391 of optically selected QSOs at $0.5<z<2.65$ with $\nu L_{\nu, 5100}$=44.7-46.5, binned in 74 luminosity-redshift bins. 
Our conclusions are:
\begin{itemize}
\item At any given redshift, an increase in the average AGN power translates to a change in the average SED shape from displaying a prominent far-IR bump to converging onto the intrinsic AGN SED of S16. This implies that the AGN contribution to the total infrared luminosity of galaxies, increases as a function of AGN power. 
\item There is no apparent correlation between SFR and AGN power, over the entire observed range in AGN power and at least at up to $z\sim3$. The SFRs of AGN host galaxies are only a function of redshift and they range from a few M$_{\odot}$/yr at $z\sim0$ to a plateau of bit less than 200\,M$_{\odot}$/yr at $z\sim2.6$, consistent with the increase in SFR density as a function of redshift. 
\item At high AGN luminosities, the total IR emission does not trace the SFR, instead it traces the AGN power. 
\item Observed trends between AGN power (measured in the optical or X-rays) and total IR emission (from the host+AGN system) can be explained by the sum of two components: (A) the infrared emission from star-formation, unrelated to AGN power and (B) the infrared emission from AGN correlated with AGN power in the optical or X-rays. At low AGN power, the AGN does not contribute significantly to the far-IR/submm hence $L_{\rm IR}$ is dominated by $L_{\rm IR, SF}$. On the other hand, in the high AGN power regime, the AGN contribution to the far-IR/submm increases and $L_{\rm IR}$ becomes dominated by $L_{\rm IR, AGN}$. As a result, the flat and subsequently increasing trend seen in the $L_{\rm AGN}$-$L_{\rm IR}$ relation, turns into an entirely flat $L_{\rm AGN}$-$L_{\rm IR, SF}$ relation once the AGN contribution is properly accounted for.
\end{itemize}

\clearpage
\section*{Acknowledgments}
MS and MJP acknowledge support by the Science and Technology Facilities Council [ST/S000216/1]. NM acknowledges support of the LMU Faculty of Physics. E.I.\ acknowledges partial support from FONDECYT through grant N$^\circ$\,1171710. M.J.M.~acknowledges the support of the National Science Centre, Poland through the SONATA BIS grant 2018/30/E/ST9/00208. JGN acknowledge financial support from the PGC 2018 project PGC2018-101948-B-I00 (MICINN, FEDER), PAPI-19-EMERG-11 (Universidad de Oviedo) and for a "Ramon y Cajal" fellowship (RYC-2013-13256) from the Spanish MINECO .

\section*{Data Availability}
The data underlying this article are available in tables both in the paper itself and online.

\bibliographystyle{mn2e}
\bibliography{references}

\clearpage
\appendix 
{\bf{APPENDIX A: The table of average SPIRE flux densities in each $L-z$ bin shown in Fig \ref{fig:bins}}}
\label{appendixA}
\vspace{10mm}

\begin{table*}
\centering
  \caption{Information for the bins shown in Fig \ref{fig:bins}. The columns are: bin number, number of objects in bin, the central redshift of the bin, the mean $\nu L_{\nu,2500}$ and the mean SPIRE flux densities with1$\sigma$ bootstrap errors }
\begin{tabular}{lcccccc} 
\hline
Bin number & number of objects & bin redshift& mean $\nu L_{\nu,2500}$ &mean $f_{250}$ & mean $f_{350}$ & mean $f_{500}$ \\
&& &log (erg/s) & mJy &mJy& mJy\\
\hline
       1&        68&  0.6&  44.79&  12.41$_{2.63}^{3.10}$&  7.99$_{1.71}^{2.23}$&  4.64$_{1.22}^{1.32}$\\
       2&        40&  0.6&  45.00&  12.66$_{1.74}^{1.86}$&  5.42$_{1.27}^{1.38}$&  -0.32$_{1.24}^{1.31}$\\
       3&        20&  0.6&  45.20&  11.41$_{2.46}^{2.24}$&  5.91$_{1.85}^{2.13}$&  0.77$_{1.93}^{1.72}$\\
       4&       117&  0.8&  44.81&  10.26$_{1.46}^{1.41}$&  6.60$_{1.21}^{1.13}$&  2.71$_{0.89}^{0.78}$\\
       5&        84&  0.8&  45.00&  13.39$_{2.47}^{2.02}$&  8.99$_{1.76}^{1.70}$&  4.27$_{1.38}^{1.25}$\\
       6&        58&  0.8&  45.19&  15.69$_{2.78}^{2.53}$&  10.70$_{2.09}^{1.92}$&  6.40$_{1.60}^{1.52}$\\
       7&        26&  0.8&  45.39&  14.69$_{2.33}^{2.17}$&  8.39$_{2.03}^{2.68}$&  2.28$_{1.83}^{2.26}$\\
       8&        18&  0.8&  45.57&  10.55$_{1.76}^{3.10}$&  5.24$_{1.76}^{2.42}$&  0.61$_{1.50}^{2.03}$\\
       9&       109&  1.0&  44.81&  8.89$_{1.52}^{1.44}$&  6.54$_{1.16}^{1.17}$&  1.91$_{0.75}^{0.92}$\\
      10&       104&  1.0&  45.00&  9.61$_{1.26}^{1.27}$&  7.42$_{1.23}^{1.23}$&  3.16$_{1.05}^{1.01}$\\
      11&        74&  1.0&  45.20&  11.23$_{1.55}^{1.71}$&  9.01$_{1.51}^{1.32}$&  5.54$_{1.25}^{1.11}$\\
      12&        74&  1.0&  45.40&  12.91$_{1.66}^{1.72}$&  10.04$_{1.23}^{1.49}$&  4.14$_{1.05}^{1.05}$\\
      13&        41&  1.0&  45.60&  14.62$_{2.31}^{2.83}$&  12.23$_{1.74}^{2.32}$&  7.24$_{1.60}^{1.38}$\\
      14&        15&  1.0&  45.78&  16.84$_{3.19}^{5.93}$&  11.53$_{3.19}^{4.50}$&  5.29$_{3.46}^{3.58}$\\
      15&        75&  1.2&  44.81&  5.95$_{1.01}^{1.00}$&  4.54$_{1.06}^{1.07}$&  2.73$_{1.00}^{1.03}$\\
      16&        86&  1.2&  45.01&  8.57$_{1.27}^{1.27}$&  5.50$_{1.11}^{1.08}$&  3.63$_{0.96}^{0.86}$\\
      17&       123&  1.2&  45.20&  8.73$_{1.02}^{1.28}$&  7.79$_{0.85}^{1.13}$&  4.21$_{0.70}^{1.06}$\\
      18&       101&  1.2&  45.40&  10.49$_{1.32}^{1.04}$&  7.73$_{1.11}^{0.89}$&  4.10$_{0.92}^{0.82}$\\
      19&        66&  1.2&  45.59&  10.50$_{1.24}^{1.36}$&  8.42$_{1.52}^{1.57}$&  4.77$_{1.47}^{1.42}$\\
      20&        43&  1.2&  45.80&  15.22$_{1.96}^{2.40}$&  9.65$_{1.75}^{2.34}$&  5.93$_{1.37}^{1.47}$\\
      21&        20&  1.2&  45.98&  16.86$_{2.90}^{3.18}$&  12.47$_{2.86}^{3.25}$&  3.83$_{1.47}^{2.18}$\\
      22&        57&  1.4&  44.80&  8.70$_{2.02}^{0.99}$&  9.39$_{1.90}^{1.38}$&  7.06$_{1.57}^{1.17}$\\
      23&        79&  1.4&  45.01&  5.90$_{1.22}^{1.25}$&  6.01$_{1.07}^{1.10}$&  4.75$_{0.99}^{1.08}$\\
      24&       135&  1.4&  45.21&  6.89$_{1.00}^{0.94}$&  6.85$_{0.88}^{0.91}$&  4.31$_{0.76}^{0.84}$\\
      25&       120&  1.4&  45.40&  11.69$_{1.30}^{1.37}$&  10.07$_{1.10}^{1.30}$&  6.44$_{1.15}^{0.97}$\\
      26&        93&  1.4&  45.59&  10.72$_{1.31}^{1.19}$&  8.83$_{1.27}^{1.06}$&  5.16$_{1.04}^{0.82}$\\
      27&        49&  1.4&  45.79&  14.49$_{2.29}^{2.56}$&  9.09$_{1.53}^{1.83}$&  4.97$_{1.22}^{1.36}$\\
      28&        20&  1.4&  45.99&  24.88$_{5.79}^{3.59}$&  20.60$_{4.19}^{3.50}$&  8.72$_{2.95}^{2.34}$\\
      29&        71&  1.6&  44.82&  7.02$_{1.19}^{1.43}$&  6.35$_{1.20}^{1.40}$&  3.48$_{1.20}^{1.37}$\\
      30&        81&  1.6&  45.02&  7.29$_{1.16}^{1.43}$&  7.63$_{0.91}^{1.13}$&  3.15$_{1.12}^{1.18}$\\
      31&        99&  1.6&  45.21&  8.89$_{1.09}^{1.23}$&  7.41$_{1.17}^{1.23}$&  5.48$_{1.07}^{1.11}$\\
      32&       121&  1.6&  45.40&  8.90$_{1.10}^{1.13}$&  6.74$_{0.88}^{0.96}$&  4.87$_{0.87}^{0.99}$\\
      33&        96&  1.6&  45.61&  9.20$_{1.10}^{1.03}$&  8.18$_{1.07}^{1.18}$&  5.71$_{0.97}^{1.11}$\\
      34&        84&  1.6&  45.80&  12.48$_{1.91}^{1.18}$&  11.34$_{1.77}^{1.30}$&  7.40$_{1.35}^{1.08}$\\
      35&        55&  1.6&  45.98&  12.86$_{1.88}^{2.19}$&  9.59$_{1.45}^{1.84}$&  6.39$_{1.52}^{1.49}$\\
      36&        32&  1.8&  44.84&  5.45$_{1.76}^{1.71}$&  4.24$_{2.15}^{1.38}$&  3.86$_{1.50}^{1.06}$\\
      37&        67&  1.8&  45.01&  6.49$_{1.14}^{1.15}$&  6.54$_{1.14}^{1.06}$&  4.43$_{1.06}^{1.21}$\\
      38&       106&  1.8&  45.21&  7.35$_{1.25}^{1.06}$&  7.29$_{1.30}^{1.13}$&  3.64$_{1.22}^{0.98}$\\
      39&       117&  1.8&  45.42&  7.06$_{1.10}^{1.11}$&  6.41$_{0.96}^{1.16}$&  4.48$_{0.95}^{0.90}$\\
      40&       112&  1.8&  45.60&  9.48$_{1.60}^{1.20}$&  8.58$_{1.77}^{1.17}$&  6.22$_{1.46}^{0.97}$\\
      41&       100&  1.8&  45.80&  12.04$_{1.58}^{1.29}$&  11.57$_{1.35}^{1.28}$&  8.19$_{1.22}^{1.12}$\\
      42&        56&  1.8&  46.00&  11.40$_{1.74}^{1.85}$&  9.40$_{1.59}^{1.83}$&  5.72$_{1.51}^{1.44}$\\
      43&        23&  1.8&  46.19&  14.11$_{2.56}^{2.54}$&  13.53$_{2.88}^{3.21}$&  4.97$_{1.74}^{2.23}$\\
      44&        55&  2.0&  45.02&  6.26$_{1.36}^{1.44}$&  6.71$_{1.33}^{1.57}$&  3.71$_{1.25}^{1.50}$\\
      45&        79&  2.0&  45.21&  7.98$_{1.55}^{1.31}$&  6.92$_{1.45}^{1.26}$&  5.58$_{1.39}^{1.16}$\\
      46&        86&  2.0&  45.41&  9.56$_{1.52}^{1.88}$&  11.31$_{1.67}^{2.06}$&  10.42$_{1.65}^{1.60}$\\
      47&       106&  2.0&  45.60&  9.26$_{1.28}^{1.22}$&  9.04$_{1.44}^{1.27}$&  6.40$_{1.14}^{1.16}$\\
      48&       101&  2.0&  45.80&  11.09$_{1.53}^{1.26}$&  8.47$_{1.48}^{1.27}$&  5.77$_{1.23}^{1.25}$\\
      49&        71&  2.0&  45.99&  11.19$_{1.78}^{1.37}$&  9.64$_{1.72}^{1.45}$&  8.25$_{1.40}^{1.37}$\\
      50&        35&  2.0&  46.18&  17.25$_{2.40}^{2.66}$&  12.72$_{1.92}^{2.43}$&  7.21$_{1.84}^{2.20}$\\
         \hline
      \end{tabular}
      \label{tab:flux}
\end{table*}

\begin{table*}
\centering
  \caption{TABLE CONTINUING --- Information for the bins shown in Fig \ref{fig:bins}. The columns are: bin number, number of objects in bin, the central redshift of the bin, the mean $\nu L_{\nu,2500}$ and the mean SPIRE flux densities with1$\sigma$ bootstrap errors }
\begin{tabular}{lcccccc} 
\hline
Bin number & number of objects & bin redshift& mean $\nu L_{\nu,2500}$ &mean $f_{250}$ & mean $f_{350}$ & mean $f_{500}$ \\
&& &log (erg/s) & mJy &mJy& mJy\\
\hline
      51&        16&  2.0&  46.39&  23.94$_{3.31}^{4.15}$&  17.38$_{3.16}^{3.87}$&  13.70$_{2.93}^{3.52}$\\
      52&        16&  2.2&  44.83&  6.55$_{2.67}^{3.85}$&  5.87$_{3.08}^{3.79}$&  7.06$_{1.74}^{3.44}$\\
      53&        81&  2.2&  45.03&  3.69$_{0.91}^{1.03}$&  4.42$_{1.23}^{1.38}$&  4.53$_{1.27}^{1.23}$\\
      54&       121&  2.2&  45.21&  6.03$_{0.90}^{0.96}$&  5.34$_{0.92}^{1.01}$&  5.70$_{0.96}^{1.07}$\\
      55&       147&  2.2&  45.42&  6.63$_{0.79}^{0.86}$&  5.96$_{0.87}^{0.90}$&  4.17$_{0.79}^{0.82}$\\
      56&       154&  2.2&  45.60&  8.52$_{0.99}^{0.98}$&  8.06$_{1.08}^{1.07}$&  5.95$_{1.01}^{1.04}$\\
      57&       111&  2.2&  45.80&  11.10$_{1.25}^{1.31}$&  9.53$_{1.30}^{1.32}$&  6.77$_{1.16}^{1.21}$\\
      58&        72&  2.2&  45.99&  9.85$_{1.39}^{1.05}$&  8.71$_{1.30}^{1.20}$&  3.55$_{1.10}^{0.84}$\\
      59&        37&  2.2&  46.19&  13.96$_{2.48}^{2.14}$&  14.85$_{2.95}^{3.11}$&  10.89$_{3.96}^{4.56}$\\
      60&        56&  2.4&  45.04&  3.07$_{1.26}^{1.41}$&  5.01$_{1.38}^{1.29}$&  3.57$_{1.47}^{1.15}$\\
      61&       109&  2.4&  45.20&  3.67$_{0.82}^{0.84}$&  2.54$_{0.91}^{0.87}$&  3.38$_{0.94}^{0.81}$\\
      62&       132&  2.4&  45.40&  4.45$_{0.77}^{0.78}$&  4.64$_{0.80}^{0.73}$&  3.59$_{0.89}^{0.87}$\\
      63&       128&  2.4&  45.61&  5.04$_{0.76}^{0.93}$&  6.08$_{0.92}^{0.90}$&  4.82$_{0.98}^{0.83}$\\
      64&       106&  2.4&  45.81&  6.93$_{1.07}^{1.01}$&  8.31$_{1.10}^{1.16}$&  6.05$_{1.12}^{1.14}$\\
      65&        76&  2.4&  45.99&  9.86$_{1.09}^{1.16}$&  8.71$_{1.34}^{1.53}$&  6.35$_{0.98}^{1.41}$\\
      66&        28&  2.4&  46.20&  13.23$_{2.77}^{1.87}$&  11.92$_{2.51}^{2.02}$&  9.33$_{2.91}^{2.14}$\\
      67&        15&  2.4&  46.41&  19.27$_{4.38}^{6.03}$&  18.26$_{3.95}^{4.47}$&  12.49$_{3.13}^{3.93}$\\
      68&        33&  2.6&  45.22&  1.74$_{1.41}^{1.03}$&  -0.47$_{1.54}^{1.29}$&  -0.26$_{1.08}^{0.92}$\\
      69&        56&  2.6&  45.41&  4.61$_{1.15}^{1.26}$&  3.35$_{1.23}^{1.26}$&  2.99$_{1.68}^{1.56}$\\
      70&        74&  2.6&  45.62&  5.71$_{1.12}^{1.15}$&  5.64$_{1.05}^{1.30}$&  3.09$_{1.19}^{1.32}$\\
      71&        65&  2.6&  45.81&  8.46$_{1.47}^{1.40}$&  8.80$_{1.79}^{1.90}$&  7.11$_{1.60}^{1.61}$\\
      72&        42&  2.6&  45.99&  5.27$_{1.09}^{0.80}$&  6.36$_{1.15}^{1.22}$&  3.36$_{1.71}^{0.97}$\\
      73&        30&  2.6&  46.18&  13.75$_{2.32}^{2.18}$&  13.48$_{2.46}^{2.35}$&  9.71$_{2.55}^{1.99}$\\
      74&        18&  2.6&  46.39&  15.33$_{3.48}^{3.92}$&  12.68$_{3.49}^{3.86}$&  10.60$_{2.50}^{2.86}$\\
        \hline
      \end{tabular}
      \label{tab:flux}
\end{table*}

\clearpage

{\bf{APPENDIX B: The data shown in Fig \ref{fig:AGNSFR_separate}}}
\label{appendixB}
\vspace{10mm}

\begin{table*}
\centering
  \caption{The data shown in Fig \ref{fig:AGNSFR_separate}. Column (1): the central redshift of the bin, Column (2): the mean $\nu L_{\nu,5100}$, Columns (3), (4) and (5): the QSO $L_{\rm IR}$ (8-1000$\mu$m, uncorrected for the AGN contribution) and lower and upper 1$\sigma$ values, Columns (6), (7) and (8): the SF $L_{\rm IR}$ (8-1000$\mu$m, AGN contribution removed) and lower and upper 1$\sigma$ values. All units are erg/s and all values are log, apart from one which is negative. }
\begin{tabular}{lccccccc} 
\hline
bin redshift& mean $\nu L_{\nu,5100}$ & $L_{\rm IR, QSO}$ & lower $L_{\rm IR, QSO}$ & upper $L_{\rm IR, QSO}$ & $L_{\rm IR, SF}$ & lower $L_{\rm IR, SF}$ & upper $L_{\rm IR, SF}$  \\
\hline
0.6 &44.66 &45.22 &45.15 &45.29 &44.99 &44.87 &45.09\\
0.6 &44.84 &45.27 &45.21 &45.32 &44.91 &44.81 &44.99\\
0.6 &44.99 &45.35 &45.28 &45.42 &44.88 &44.67 &45.02\\
0.8 &44.70 &45.29 &45.23 &45.34 &45.07 &44.98 &45.14\\
0.8 &44.81 &45.42 &45.35 &45.47 &45.21 &45.10 &45.29\\
0.8 &44.99 &45.52 &45.45 &45.57 &45.25 &45.14 &45.33\\
0.8 &45.18 &45.56 &45.49 &45.61 &45.12 &44.98 &45.21\\
0.8 &45.36 &45.61 &45.53 &45.68 &44.76 &44.26 &45.04\\
1.0 &44.73 &45.33 &45.26 &45.38 &45.11 &45.02 &45.19\\
1.0 &44.86 &45.42 &45.36 &45.46 &45.17 &45.10 &45.24\\
1.0 &45.05 &45.53 &45.47 &45.59 &45.23 &45.14 &45.32\\
1.0 &45.25 &45.64 &45.58 &45.70 &45.22 &45.11 &45.31\\
1.0 &45.38 &45.74 &45.67 &45.81 &45.25 &45.09 &45.39\\
1.0 &45.51 &45.82 &45.75 &45.90 &45.22 &45.00 &45.44\\
1.2 &44.71 &45.30 &45.23 &45.36 &45.08 &44.97 &45.16\\
1.2 &44.81 &45.45 &45.39 &45.50 &45.26 &45.18 &45.33\\
1.2 &44.99 &45.51 &45.45 &45.56 &45.23 &45.16 &45.31\\
1.2 &45.13 &45.61 &45.55 &45.65 &45.30 &45.22 &45.36\\
1.2 &45.36 &45.71 &45.65 &45.77 &45.21 &45.09 &45.31\\
1.2 &45.56 &45.88 &45.82 &45.94 &45.32 &45.16 &45.44\\
1.2 &45.70 &45.99 &45.92 &46.06 &45.33 &45.09 &45.49\\
1.4 &44.72 &45.51 &45.41 &45.55 &45.38 &45.26 &45.44\\
1.4 &44.81 &45.41 &45.33 &45.48 &45.20 &45.08 &45.30\\
1.4 &45.02 &45.50 &45.44 &45.55 &45.20 &45.10 &45.28\\
1.4 &45.17 &45.71 &45.66 &45.76 &45.46 &45.39 &45.53\\
1.4 &45.33 &45.74 &45.68 &45.79 &45.34 &45.24 &45.42\\
1.4 &45.52 &45.90 &45.83 &45.96 &45.46 &45.32 &45.58\\
1.4 &45.69 &46.11 &46.03 &46.17 &45.73 &45.56 &45.82\\
1.6 &44.77 &45.51 &45.44 &45.58 &45.37 &45.28 &45.46\\
1.6 &44.89 &45.54 &45.48 &45.61 &45.36 &45.27 &45.45\\
1.6 &45.04 &45.65 &45.60 &45.70 &45.44 &45.37 &45.51\\
1.6 &45.21 &45.70 &45.65 &45.75 &45.41 &45.32 &45.48\\
1.6 &45.42 &45.79 &45.73 &45.85 &45.33 &45.22 &45.41\\
1.6 &45.59 &45.94 &45.88 &45.99 &45.46 &45.31 &45.53\\
1.6 &45.76 &46.04 &45.97 &46.11 &45.33 &45.09 &45.50\\
1.8 &44.80 &45.49 &45.36 &45.59 &45.32 &45.12 &45.46\\
1.8 &44.84 &45.55 &45.48 &45.61 &45.39 &45.29 &45.47\\
1.8 &45.00 &45.65 &45.58 &45.70 &45.46 &45.36 &45.53\\
1.8 &45.19 &45.68 &45.62 &45.74 &45.38 &45.28 &45.47\\
1.8 &45.35 &45.83 &45.76 &45.88 &45.51 &45.40 &45.58\\
1.8 &45.52 &45.95 &45.89 &46.00 &45.59 &45.49 &45.66\\
1.8 &45.73 &46.04 &45.97 &46.10 &45.43 &45.24 &45.57\\
1.8 &45.86 &46.16 &46.08 &46.22 &45.49 &45.23 &45.65\\
2.0 &44.88 &45.62 &45.54 &45.69 &45.48 &45.36 &45.57\\
2.0 &45.01 &45.73 &45.66 &45.79 &45.58 &45.48 &45.66\\
2.0 &45.16 &45.82 &45.75 &45.89 &45.64 &45.55 &45.73\\
2.0 &45.31 &45.85 &45.80 &45.90 &45.60 &45.51 &45.67\\
2.0 &45.52 &45.98 &45.92 &46.03 &45.66 &45.56 &45.73\\
2.0 &45.70 &46.05 &45.98 &46.11 &45.56 &45.40 &45.65\\
2.0 &45.92 &46.26 &46.19 &46.32 &45.74 &45.59 &45.85\\
2.0 &46.10 &46.42 &46.36 &46.48 &45.85 &45.70 &45.98\\
       \hline
      \end{tabular}
      \label{tab:datafig3}
\end{table*}

\begin{table*}
\centering
  \caption{TABLE CONTINUING - The data shown in Fig \ref{fig:AGNSFR_separate}. Column (1): the central redshift of the bin, Column (2): the mean $\nu L_{\nu,5100}$, Columns (3), (4) and (5): the QSO $L_{\rm IR}$ (8-1000$\mu$m, uncorrected for the AGN contribution) and lower and upper 1$\sigma$ values, Columns (6), (7) and (8): the SF $L_{\rm IR}$ (8-1000$\mu$m, AGN contribution removed) and lower and upper 1$\sigma$ values. All units are erg/s and all values are log, apart from one which is negative. }
\begin{tabular}{lccccccc} 
\hline
bin redshift& mean $\nu L_{\nu,5100}$ & $L_{\rm IR, QSO}$ & lower $L_{\rm IR, QSO}$ & upper $L_{\rm IR, QSO}$ & $L_{\rm IR, SF}$ & lower $L_{\rm IR, SF}$ & upper $L_{\rm IR, SF}$  \\
\hline
2.2 &44.84 &45.69 &45.51 &45.87 &45.59 &45.33 &45.81\\
2.2 &44.94 &45.53 &45.43 &45.61 &45.30 &45.14 &45.43\\
2.2 &45.05 &45.71 &45.65 &45.77 &45.53 &45.45 &45.61\\
2.2 &45.21 &45.78 &45.73 &45.83 &45.55 &45.48 &45.62\\
2.2 &45.37 &45.91 &45.85 &45.95 &45.65 &45.58 &45.71\\
2.2 &45.56 &46.05 &45.99 &46.10 &45.75 &45.67 &45.81\\
2.2 &45.73 &46.08 &46.02 &46.14 &45.58 &45.44 &45.67\\
2.2 &45.95 &46.27 &46.20 &46.33 &45.68 &45.47 &45.81\\
2.4 &44.98 &45.53 &45.37 &45.65 &45.27 &44.96 &45.47\\
2.4 &45.07 &45.61 &45.53 &45.68 &45.36 &45.21 &45.47\\
2.4 &45.25 &45.73 &45.66 &45.79 &45.42 &45.30 &45.51\\
2.4 &45.39 &45.81 &45.74 &45.87 &45.43 &45.31 &45.53\\
2.4 &45.60 &45.99 &45.92 &46.04 &45.54 &45.40 &45.64\\
2.4 &45.75 &46.13 &46.07 &46.19 &45.70 &45.60 &45.78\\
2.4 &46.00 &46.32 &46.24 &46.38 &45.73 &45.48 &45.85\\
2.4 &46.16 &46.48 &46.40 &46.56 &45.91 &45.62 &46.12\\
2.6 &45.08 &45.45 &45.14 &45.60 &45.00 &-4.38$\times$10$^{44}$ &45.31\\
2.6 &45.21 &45.77 &45.67 &45.85 &45.53 &45.36 &45.66\\
2.6 &45.37 &45.88 &45.80 &45.94 &45.60 &45.47 &45.70\\
2.6 &45.57 &46.06 &45.99 &46.12 &45.76 &45.65 &45.85\\
2.6 &45.72 &46.01 &45.92 &46.07 &45.32 &44.98 &45.47\\
2.6 &45.96 &46.34 &46.27 &46.40 &45.89 &45.73 &45.99\\
2.6 &46.12 &46.43 &46.35 &46.50 &45.84 &45.55 &46.02\\
       \hline
      \end{tabular}
      \label{tab:datafig3}
\end{table*}

\end{document}